\documentclass[fleqn,11pt]{article}
\usepackage{hyperref}
\usepackage{amsmath, amsthm,amssymb, multirow, graphicx,footnote,caption,bm,color}
\usepackage{color}

\usepackage{apacite}

\usepackage{amsmath,bm}
\usepackage{color}
\usepackage{rotating}
\usepackage{caption}
\usepackage{subcaption}
\usepackage{grffile}
\usepackage{graphicx}
\usepackage{lscape}
\usepackage{blindtext}
\usepackage{amssymb}
\usepackage{varioref} 
\usepackage{textcomp}
\usepackage{booktabs}
\usepackage{multirow}
\usepackage{pdflscape}
\usepackage{relsize}
\usepackage{tablefootnote}
\usepackage{wrapfig}
\usepackage{natbib}
\usepackage{listings}
\usepackage{longtable}
\usepackage{url}
\usepackage{tabularx}
\usepackage{float}
\usepackage{epsfig}
\usepackage[titletoc]{appendix}
\usepackage{arydshln}
\usepackage{footnote}
\usepackage{secdot}
\usepackage{bigints}
\usepackage[utf8]{inputenc}
\usepackage[english]{babel}
\usepackage{xr}
\usepackage{adjustbox}
\usepackage{graphicx} 

\usepackage{multicol}
\usepackage{enumitem}
\usepackage{bigstrut}
\usepackage{rotating}
\usepackage{xr}

\begin{document}

\title{\bf{Likelihood-Based Inference for Semi-Parametric Transformation Cure Models with Interval Censored Data}}
\author{\bf{Suvra Pal${}^{1}$ and Sandip Barui${}^{2,\footnote{Corresponding author. E-mail address: sandipbarui@iimk.ac.in;  Tel.: +(91)-495-2809-684.\newline \indent }}$}  \\\\
${}^{1}$Department of Mathematics, University of Texas at Arlington, \\
411 S Nedderman Drive, Arlington, TX, 76019, USA. \\
${}^{2}$Quantitative Methods and Operations Management Area, \\
Indian Institute of Management Kozhikode, Kozhikode, Kerala, 673570, India.
}
\date{}

\maketitle

\begin{abstract}
A simple yet effective way of modeling survival data with cure fraction is by considering Box-Cox transformation cure model (BCTM) that unifies mixture and promotion time cure models. In this article, we numerically study the statistical properties of the BCTM when applied to interval censored data. Time-to-events associated with susceptible subjects are modeled through proportional hazards structure that allows for non-homogeneity across subjects, where the baseline hazard function is estimated by distribution-free piecewise linear function with varied degrees of non-parametricity. Due to missing cured statuses for right censored subjects, maximum likelihood estimates of model parameters are obtained by developing an expectation-maximization (EM) algorithm.  Under the EM framework, the conditional expectation of the complete data log-likelihood function is maximized by considering all parameters (including the Box-Cox transformation parameter $\alpha$) simultaneously, in contrast to conventional profile-likelihood technique of estimating $\alpha$. The robustness and accuracy of the model and estimation method are established through a detailed simulation study under various parameter settings, and an analysis of real-life data obtained from a smoking cessation study.        
\end{abstract}



\section{Introduction}
Over recent years, cure rate survival models or cure models have gained much popularity in both medical and statistical fields (\citealp{othus2012cure,peng2014cure,Pal18,Majak19,felizzi2021mixture}). With advent of revolutionary treatment techniques in the drug development industry, chances of patients getting cured from various critical and life-threatening conditions have gone higher. In contrast to regular survival models which rely on the assumption that all subjects under study will inevitably face the event of interest (death, relapse etc.) when observed long enough \citep{Tom21,Hoang23}, cure models incorporate a cure probability (or, cure rate) to account for the proportion of subjects who may not face the event of interest at all. The notion was introduced in the works of \cite{Boa49}, and later theorized and validated under a formal model framework by \cite{Ber52}. In general, cure models assume the population to be a mixture of susceptible subjects (or, non-cured) who will eventually face the event and immune subjects (or, cured) who will not face the event even when they are followed-up for a sufficiently long duration of time. For well-designed prospective studies with higher efficacy of novel therapeutic agents or treatments for critical conditions, the end-of-study survival plot (e.g., the Kaplan Meier plot) is likely to level-off to a value significantly higher than zero after showing a plateau, indicating the existence of cure rate (\citealp{Sy00}).

A popular way of statistically modeling survival data that indicate the presence of a cure component is by considering  time-to-event random variable $Y = (1-\eta) Y_{c}+ \eta Y_{s}$, where $\eta=1$ if the subject is susceptible and $\eta=0$ if the subject is cured. Therefore, if $\eta=1$, $Y=Y_s$ is the susceptible time-to-event, and if $\eta=0$, $Y=Y_c=\infty$ is the cured time-to-event. Note that, $P(\eta \notin \{0, 1\})=0$ and $\pi=P(\eta=0)$ is the cure probability whose estimate is of primary interest. If $S(\cdot)$ denotes the survival function corresponding to susceptible subject, the resultant unconditional survival function (population survival function) of any subject in the population can be expressed as
\begin{equation}
    S_p(y)=\pi + (1-\pi)S(y).
    \label{mix}
\end{equation}
Here, $\pi$ and $S(\cdot)$ are called the incidence and latency parts of a cure model, respectively. Also, note that $\pi=P(\eta=0)=\lim_{y \to \infty}S_p(y)$. The cure model in \eqref{mix}, known as the mixture cure model (MCM), has been widely studied under various framework and considerations in survival analysis literature. For example, Farewell studied the MCM by assuming a Weibull distribution for time-to-events corresponding to the susceptible individuals, and linked covariates to the cured probability through logistic-link function (\citealp{Far82}). Kuk and Chen linked covariates to the susceptible time-to-events by introducing proportional hazards (PH) structure where a marginal-likelihood approach with expectation-maximization (EM) algorithm was applied to estimate baseline hazard function and regression parameters (\citealp{Kuk92}). By considering a model structure similar to Kuk and Chen, Sy and Taylor studied a Breslow-type estimator while applying EM algorithm to jointly estimate the incidence and latency parameters (\citealp{Sy00, Sy01}). The properties of the MCM as a special case of some flexible cure models have been investigated in details by \cite{Bal12, Bal13, Bal15, Bal16b}.  
Traditionally, a logistic-link function has been applied to link covariates to the incidence part (\citealp{Far82,Bal12, Bal15,balakrishnan2017proportional,Bala-Barui22,balakrishnan2023destructive}). More recently, it has been shown by \cite{Paletal23} and \cite{Ase23} that the predictive accuracy of the cure probability can be improved by applying suitable machine learning algorithm. The MCM has also been examined under a semi-parametric set-up by considering additive hazards model with study covariates prone to measurement error (\citealp{barui2020semiparametric}).

A more realistic approach to modeling time-to-event data in survival studies marked by cured proportion is to consider a competing causes scenario where it is assumed that an unknown number of competing risk factors affect time-to-events corresponding to susceptible subjects \cite{Pal16,Pal17a,Pal17b}. An estimate of the probability of a subject whose time-to-event is not influenced by any competing causes is considered as the cure rate. This kind of modeling technique has been studied by \cite{Yak96,Ibr01,Tso03,Rod09, Bal16, balakrishnan2017proportional, Xie21a} and \cite{PalSVM23}, to name a few. Considering $M$ as the unobserved number of risk factors competing to produce an event, where $M$ is assumed to follow a Poisson distribution with mean $\phi>0$, gives rise to the promotion-time cure model (PCM) with population survival function given by
\begin{equation}
    S_p(y)= \exp\{-\phi (1-S(y))\}.
\end{equation}
In this case, the cured probability is expressed as $\pi=\lim_{y \to \infty}S_p(y)=\exp\{-\phi\}$, where $S(\cdot)$ denotes the common survival function of the time-to-events corresponding to $M$ competing causes.

A convenient and flexible way to unify the MCM and PCM under a single class of models is by considering a Box-Cox transformation cure model (BCTM) where the Box-Cox transformation is applied on $S_p(y)$ with transformation parameter $\alpha$. The BCTM was introduced by \cite{Yin05} and later studied by \cite{Zen06,Kou17,Pal18c,Mil22,Wang22} and \cite{PalRoy23} under various assumptions and model settings, especially for right censored data. However, based on a detailed and systematic literature search, we did not find any published study that deals specifically with the application of BCTM when the observed data is non-informative interval censored in nature. Analysis of interval censored data under cure model set-up, in general, has been carried out by few researchers. For example,  \cite{Kim08,Ma10,Xia11,Pal17c,Wia18,Jodi22,JodiSIM23} considered fully parametric, semi-parametric and distribution-free frameworks. For survival studies where subjects are monitored or followed-up at scheduled discrete time points during the study period, interval censored data are obtained. For subjects who experience the event of interest, the time interval (with finite limits) in which the event occurs is observed and recorded. For subjects who do not experience the event within the study period, the last visit time is considered as the left limit of the corresponding time interval while the right limit is taken as infinite. Therefore, in this article, we carry out an in-depth analysis of the BCTM in the presence of non-informative interval censored data. 

In our proposed model, the survival function $S(\cdot)$ of the lifetime corresponding to the latency part is modeled by assuming the corresponding hazard function $\lambda(\cdot)$ to have a PH structure, where the baseline hazard function $\lambda_0(\cdot)$ is estimated by piecewise linear approximation (PLA) in a distribution-free or non-parametric way \citep{Bala-Barui22}. This is a convenient and robust way of modeling time-to-events associated with the susceptible subjects without making any strong assumptions about their lifetime distribution. The PH structure allows us to account for heterogeneity that is present across subjects with respect to covariates. In the context of PH structure, the baseline hazard function has been estimated by piecewise constant function (\citealp{Lar85}) and PLA (\citealp{Lo93}). For the PCM, $S(\cdot)$ has been approximated by piecewise constant function (\citealp{Che01}). Some semi-parametric cure models, e.g., mixture, promotion-time, geometric and Conway-Maxwell (COM) Poisson cure models have been studied where the baseline hazard function is approximated by PLA (\citealp{Bal16b, Bala-Barui22}). A comparison of the model fit and estimation saccuracy between fully parametric MCM with $S(\cdot)$ corresponding to an exponentiated Weibull distribution and semi-parametric MCM with $\lambda_0(\cdot)$ estimated by PLA has also been studied recently (\citealp{pal2022stochastic}). Typically, piecewise linear function in PLA is characterized by the number of cut-points or knots ($B \in \mathbb{Z}^+$) and choices of the cut-points denoted by $\tau_0 <, \dots, <\tau_B$, both of which are user-defined. Similar to smoothing splines, $B$ represents the degree of non-parametricity of the model, meaning models with higher $B$ exhibit greater extent of distribution-free features than models with lower $B$ (\citealp{Bal16b}). However, models with higher $B$ suffer from over-fitting and estimation complications associated with models having many parameters. The choices of $\tau_b, b=0, \dots, B$, should be made logically to estimate $\lambda_0(\cdot)$ with reasonable accuracy. Therefore, our main contributions to the area of cure models, through this article, can be described by the following: For interval censored data, through a detailed simulation study and a real-life example, the behavior of the BCTM is examined numerically by assuming that the hazard function associated with the latency part follows a proportional hazards structure, and the baseline hazard function is estimated by piecewise linear function in a distribution-free manner. The model parameters of the resulting semi-parametric BCTM are estimated by the usual maximum likelihood (ML) estimation by incorporating the EM algorithm. Most literature implements a profile-likelihood technique on the transformation parameter $\alpha$ of the BCTM (\citealp{YinIb05, Yin05b,Pal18c}). However, we carry out the estimation by maximizing the conditional expectation of the log-likelihood function with respect to the all parameters simultaneously (also known as simultaneous-maximization technique).

The rest of the article is arranged in the following manner. In Section \ref{sec:sec2}, the theoretical framework of the proposed model is presented in detail. For interval censored data, the mathematical expressions required to implement the EM algorithm are derived in Section \ref{sec:sec3}. In Section \ref{sec:sec4}, the performance (e.g., robustness, accuracy, etc.) of the proposed model and estimation method on interval censored data is demonstrated by carrying out a simulation study under various parameter settings. Section \ref{sec:sec5} re-establishes the model performance and usefulness via an application on the interval censored data based on a real-life study on smoking cessation. Finally, we provide some concluding remarks and suggestions for future research in Section \ref{sec:sec6}.

\section{Transformation cure model framework} \label{sec:sec2}

A Box-Cox transformation with transformation parameter $\alpha$ on a variable $u$ is given by 
\begin{equation}
G(u,\alpha)=\begin{cases}
\frac{u^\alpha -1}{\alpha}, & \alpha \ne 0 \\
\ln\{u\}, & \alpha = 0.
\end{cases}
\label{BC} 
\end{equation}  
Theoretically, $\alpha$ can take any real value. However, the proposed cure model (BCTM) that unifies MCM and PCM under a single framework can be established with $\alpha \in [0,1]$. Considering $G(S_p(y), \alpha)=- \pi(\alpha) (1-S(y))$, from eqn.(\ref{BC}), we obtain the following:
\begin{equation}
S_p(y)=\begin{cases}
{\left[1-\alpha \pi(\alpha)(1-S(y))\right]^{1/\alpha}}, & 0 < \alpha \leq 1 \\
\exp\{-\pi(\alpha)(1-S(y))\}, & \alpha = 0,
\end{cases}
\label{BC2} 
\end{equation}
where $\pi(\cdot)$ is linked to the covariate vector $\bm z = (1, \bm z^{*\intercal})^{\intercal}$ of order $q_1+1$ by
\begin{equation}
\pi(\alpha)=\pi(\alpha, \bm \beta; \bm z)=\begin{cases}
\frac{\exp\left\{\bm z^{\intercal}\bm \beta\right\}}{1+ \alpha \exp\left\{\bm z^{\intercal}\bm \beta\right\}}, & 0 < \alpha \leq 1 \\
\exp\left\{\bm z^{\intercal}\bm \beta\right\}, & \alpha = 0.
\end{cases}
\label{BC3} 
\end{equation}
Here, $\pi(\cdot)$ is the part of the model that accounts for the incidence while $S(\cdot)$ accounts for the latency. Note that $\alpha$ is the parameter that regulates the structure of the cure model. The MCM and PTCM are, therefore, BCTM with $\alpha=1$ and $\alpha=0$, respectively. Therefore, combining eqns.(\ref{BC2}) and (\ref{BC3}), the population survival function for the BCTM can be expressed as
\begin{equation}
S_p(y)=S_p(y; \bm z, \alpha, \bm \beta)=\begin{cases}
{\left[1- \frac{\alpha \exp\left\{\bm z^{\intercal}\bm \beta\right\}}{1+ \alpha \exp\left\{\bm z^{\intercal}\bm \beta\right\}}(1-S(y))\right]^{1/\alpha}}, & 0 < \alpha \leq 1 \\
\exp\left\{-\exp\left\{\bm z^{\intercal}\bm \beta\right\}(1-S(y))\right\}, & \alpha = 0.
\end{cases}
\label{BC4} 
\end{equation}

The discussion on modeling the latency part is as follows. Being inspired by the works of \cite{Kou17} and \cite{Bala-Barui22}, $S(y)$ is modeled by a PH structure with the baseline hazard function $\lambda_0(y)=\lambda_0(y; \bm \psi)$ as 
\begin{equation}\label{PLAdef}
    \lambda_0(y; \bm \psi)=\sum_{b=1}^{B}\left[\psi_b+ \left\{\frac{\psi_b-\psi_{b-1}}{\tau_{b}-\tau_{b-1}}\right\} (y-\tau_b) \right] I_{[\tau_{b-1},\tau_b]}(y),
\end{equation} where \begin{equation*}
I_{[\tau_{b-1},\tau_b]}(y)=\begin{cases}
1, & y \in [\tau_{b-1},\tau_b]\\
0, & \text{otherwise}
\end{cases}
\label{Indicator}    
\end{equation*}
for $b=1, \dots, B$ and a set of user-defined cut-points $\tau_0 < \tau_1 < \dots < \tau_B$. Eqn. (\ref{PLAdef}) is considered such that $\lim_{y \to \tau_b}\lambda_0(y; \bm \psi)=\psi_b$. In other words, we approximate the baseline hazard function of any susceptible subject in the population with a piecewise linear function defined in eqn. (\ref{PLAdef}) with continuity at the cut-points in the set $\{\tau_0, \tau_1, \dots, \tau_B\}$. Further, to ensure that time-to-event for a susceptible individual varies with respect to covariate vector $\bm x$ of order $q_2$, a PH structure is considered for modeling the hazard function $\lambda(y; \bm x, \bm \psi, \bm \gamma)$ in the usual way, i.e., $\lambda(y;  \bm x,\bm \psi, \bm \gamma)=\lambda_0(y; \bm \psi)\exp\{\bm x^{\intercal} \bm \gamma\}$. Note that, $\bm z$ and $\bm x$ may share same covariate elements or may be the same. This implies that the survival function $S(y)$ can be expressed as
\begin{equation}\label{sussurv}
 S(y)=S(y; \bm x, \bm \psi,  \bm \gamma)=\exp\{-\Lambda(y;  \bm x,\bm \psi, \bm \gamma)\}   
\end{equation}
where 
\begin{flalign}
    &\Lambda(y;  \bm x,\bm \psi, \bm \gamma)=\exp\{\bm x^{\intercal} \bm \gamma\}\int_0^y \lambda_0(w; \bm \psi) dw 
\end{flalign}
is the cumulative hazard function for susceptible individual, and 
\begin{flalign}\label{BCHF}    
     &\int_0^y \lambda_0(w; \bm \psi) dw=\sum_{b=1}^B \psi_b \left[\min\{y,\tau_b\}-\tau_{b-1}\right]I_{[\tau_{b-1}, \infty)}(y)  \nonumber \\
    &+\sum_{b=1}^B\left[\left\{\frac{\psi_b-\psi_{b-1}}{\tau_{b}-\tau_{b-1}}\right\} \left\{\frac{\min\{y,\tau_b\}^2-\tau^2_{b-1}}{2}\right\} - \tau_b \left\{\min\{y,\tau_b\}-\tau_{b-1}\right\}\right] I_{[\tau_{b-1},\infty)}(y).
\end{flalign}
Finally, eqn. (\ref{BC4}) takes the following form:
\begin{equation}\label{BC5}
S_p(y)=S_p(y; \bm x, \bm z, \bm \theta)=\begin{cases}
{\left[1- \frac{\alpha \exp\left\{\bm z^{\intercal}\bm \beta\right\}}{1+ \alpha \exp\left\{\bm z^{\intercal}\bm \beta\right\}}\left(1-\exp\{-\Lambda(y;  \bm x,\bm \psi, \bm \gamma)\}\right)\right]^{1/\alpha}}, & 0 < \alpha \leq 1 \\
\exp\left\{-\exp\left\{\bm z^{\intercal}\bm \beta\right\}\left(1-\exp\{-\Lambda(y;  \bm x,\bm \psi, \bm \gamma)\}\right)\right\}, & \alpha = 0.
\end{cases}
\end{equation}
Some approaches to select the cut points in $\{\tau_0, \tau_1, \dots, \tau_B\}$ are discussed in  Sections \ref{sec:sec3} and \ref{sec:sec4}. 
The linear function obtained for the interval $[\tau_{B-1}, \tau_B]$, i.e., $\left\{\psi_B-\tau_B\left(\frac{\psi_B-\psi_{B-1}}{\tau_B-\tau_{B-1}}\right)\right\}+\left\{\left(\frac{\psi_B-\psi_{B-1}}{\tau_B-\tau_{B-1}}\right)t\right\}$ is extended to the interval $[\tau_B, \infty)$ for any $t>\tau_B$. The parameter vector of the model is $\bm \theta=(\alpha, \bm \psi, \bm \beta, \bm \gamma)^{\intercal}$ with corresponding parameter space as $\Theta \in \mathbb{D} \subset \mathbb{R}^{(B+q_1+q_2+3)}$. The cure rate of the BCTM model is simply given by
\begin{equation}\label{CR}
\pi(\bm \theta; \bm z)=\pi(\alpha, \bm \beta; \bm z)=\lim_{y \to \infty} S_p(y; \bm x, \bm z, \bm \theta)=\begin{cases}
{\left[1- \frac{\alpha \exp\left\{\bm z^{\intercal}\bm \beta\right\}}{1+ \alpha \exp\left\{\bm z^{\intercal}\bm \beta\right\}}\right]^{1/\alpha}}, & 0 < \alpha \leq 1 \\
\exp\left\{-\exp\left\{\bm z^{\intercal}\bm \beta\right\}\right\}, & \alpha = 0.
\end{cases}
\end{equation}


\section{Simultaneous-maximization under EM framework for interval censored data} \label{sec:sec3}

In this Section, we discuss the method of parameter estimation by maximum likelihood approach under the EM framework for our proposed model in the presence of interval censored data. The expected value of the complete data log-likelihood function, given the observed data and current value of the parameter estimates, is maximized with respect to the parameter vector $\bm \theta$, that is, considering all parameters simultaneously, also known as the simultaneous-maximization technique. 

In the context of right censored data, earlier studies such as \cite{YinIb05}, \cite{Yin05b} and \cite{Pal18c} have used profile-likelihood technique on $\alpha$ for the BCTM citing model identifiability as a concern which makes parameter estimates unstable with high standard errors. However, few recent studies such as \cite{Wang22} and \cite{PalRoy23} have established accuracy and consistency for the simultaneous-maximization technique for right censored data when the search of $\alpha$ is restricted to $[0,1]$. To the best of our knowledge, this article is the first one to propose and carry out simultaneous-maximization under the EM-based BCTM framework with PLA of the baseline hazard function for interval censored data. Since one of our objectives is to establish a flexible family of cure model that unifies both the MCM and PCM under the same umbrella, we focus our search for $\alpha$ only in the interval $[0,1]$.

For a sample of size $n$, the observed interval censored data can be represented as $$\mathcal{D}_O=\{(l_i, r_i,\delta_i, \bm x_i, \bm z_i): i=1, \dots, n\}.$$ For subject $i$,  $l_i$ and $r_i$ are the lower and upper limits (with $l_i<r_i$) of the interval in which the actual time-to-event (unobserved) $y_i$ belongs, and $\delta_i$ is the censoring indicator such that $\delta_i=1$ if both $l_i$ and $r_i$ take finite values (interval censored), else $\delta_i=0$ (right censored). Similarly, $\bm x_i$ and $\bm z_i$ are the observed covariate vectors corresponding to the latency and incidence parts, respectively, for subject $i$. For interval-censored data as defined above, the observed data log-likelihood function is given by
\begin{flalign}\label{lobs}
    l(\bm \theta| \mathcal{D}_O) & = \sum_{i\in\Delta_1}\ln\{ S_p(l_i;\bm x_i,\bm z_i, \bm \theta)-S_p(r_i;\bm x_i,\bm z_i, \bm \theta)\} + \sum_{i\in \Delta_0} \ln\{ S_p(l_i;\bm x_i,\bm z_i, \bm \theta)\},
\end{flalign}
where $\Delta_1=\{i: \delta_i=1\}$ and $\Delta_0=\{i: \delta_i=0\}$. However, we have partial information on the cure status $\eta_i$ for subject $i$, i.e., $\eta_i=1$ if $i \in \Delta_1$ and $\eta_i$ is missing for $i \in \Delta_0$. Therefore, the complete interval-censored data can be represented by $\mathcal{D}_C=\{(l_i, r_i,\delta_i, \bm x_i, \bm z_i, \eta_i): i=1, \dots, n\}$, which includes both observed and unobserved $\eta_i$s. An effective way to handle missing information related to $\eta_i$ for $i \in \Delta_0$ under maximum likelihood estimation set-up is through the application of the EM algorithm. For the BCTM, the complete data log-likelihood function is given by
\begin{flalign}\label{lcom}
    l(\bm \theta| \mathcal{D}_C) & = \sum_{i\in\Delta_1}\ln\{ S_p(l_i;\bm x_i,\bm z_i, \bm \theta)-S_p(r_i;\bm x_i,\bm z_i, \bm \theta)\} \\ \nonumber 
    & + \sum_{i\in \Delta_0} (1-\eta_i)\ln\{ \pi(\bm \theta; \bm z_i)\} + \sum_{i\in\Delta_0} \eta_i \ln\{ [1-\pi(\bm \theta; \bm z_i)]S_u(l_i;\bm x_i,\bm z_i, \bm \theta) \},
\end{flalign}
where $S_p(\cdot; \bm x_i, \bm z_i, \bm \theta)$ and $\pi(\bm \theta; \bm z_i)$ are given in eqns. (\ref{BC5}) and (\ref{CR}), respectively, and $S_{u}(y_i;\bm x_i,\bm z_i, \bm \theta)$ is the survival function of susceptible subject, which can be expressed as
\begin{eqnarray}
    S_{u}(y_i;\bm x_i,\bm z_i, \bm \theta) &=& \frac{S_p(y_i;\bm x_i,\bm z_i, \bm \theta)-\pi(\bm \theta; \bm z_i)}{1-\pi(\bm \theta; \bm z_i)}\\ 
    &=& \begin{cases} 
\frac{{\left[1- \frac{\alpha \exp\left\{\bm z_i^{\intercal}\bm \beta\right\}}{1+ \alpha \exp\left\{\bm z_i^{\intercal}\bm \beta\right\}}\left(1-\exp\{-\Lambda(y_i;  \bm x_i,\bm \psi, \bm \gamma)\}\right)\right]^{1/\alpha}} - {\left[1- \frac{\alpha \exp\left\{\bm z_i^{\intercal}\bm \beta\right\}}{1+ \alpha \exp\left\{\bm z_i^{\intercal}\bm \beta\right\}}\right]^{1/\alpha}}}{1-{\left[1- \frac{\alpha \exp\left\{\bm z_i^{\intercal}\bm \beta\right\}}{1+ \alpha \exp\left\{\bm z_i^{\intercal}\bm \beta\right\}}\right]^{1/\alpha}}} , & 0 < \alpha \leq 1\\
\frac{\exp\left\{-\exp\left\{\bm z_i^{\intercal}\bm \beta\right\}\left(1-\exp\{-\Lambda(y_i;  \bm x_i,\bm \psi, \bm \gamma)\}\right)\right\}-\exp\left\{-\exp\left\{\bm z_i^{\intercal}\bm \beta\right\}\right\}}{1-\exp\left\{-\exp\left\{\bm z_i^{\intercal}\bm \beta\right\}\right\}}, & \alpha = 0.\\
\end{cases}
\label{SusS}
\end{eqnarray}
Following the steps in the works of \cite{Pal17c} and \cite{Wia18}, we obtain the ML estimate of $\bm \theta$ by implementing EM algorithm to interval censored data.
\subsection{Implementation of the EM algorithm}
At the $k$-th iteration:  
\begin{enumerate}
\item[1.] {\bf E-Step}: we calculate 
\begin{flalign}\label{Q}
Q(\bm\theta,\bm\theta^{(k)}) &=\sum_{i\in\Delta_1}\ln\{ S_p(l_i;\bm x_i,\bm z_i, \bm \theta)-S_p(r_i;\bm x_i,\bm z_i, \bm \theta\} + \sum_{i\in \Delta_0} (1-w_i^{(k)})\ln\{ \pi(\bm \theta; \bm z_i)\} \nonumber\\
&+\sum_{i\in\Delta_0} w_i^{(k)} \ln\{ (1-\pi(\bm \theta; \bm z_i)) S_u(l_i;\bm x_i,\bm z_i, \bm \theta \},
\end{flalign}
where
\begin{equation}
w_i^{(k)}=E(\eta_i|T_i>l_i;\boldsymbol\theta^{(k)})=P(\eta_i=1|T_i>l_i;\boldsymbol\theta^{(k)})=\left.\frac{(1-\pi(\bm \theta; \bm z_i))S_u(l_i;\bm x_i,\bm z_i, \bm \theta)}{S_p(l_i;\bm x_i,\bm z_i, \bm \theta)}\right|_{\boldsymbol\theta=\boldsymbol\theta^{(k)}}.
\end{equation}

\item[2.] {\bf M-step}: we obtain an improved estimate of $\bm\theta$ as
\begin{equation}\hat {\bm \theta}^{(k+1)}= \underset{\bm \theta \in  \Theta} {\arg \max \hspace{1mm}} Q(\bm\theta,\bm\theta^{(k)}).
\end{equation}

\item[3.] {\bf Iterative-step}: 
The E-step and M-step are repeated until a desired level of tolerance is achieved for a suitable convergence criterion to obtain the ML estimate $\hat {\bm \theta}$ of $\bm {\theta}$. In our case, we considered the convergence criterion as $\bigg| \frac{\boldsymbol\theta^{(k+1)}-\boldsymbol\theta^{(k)}}{\boldsymbol\theta^{(k)}} \bigg| < \epsilon,$ where $\epsilon=10^{-3}$.

\end{enumerate}
The M-step of the EM algorithm is carried out in software {\tt R v.4.0.4} using Nelder-Mead or L-BFGS-B methods conveniently in routine {\tt optim} under {\tt optimx} library. Interested readers may also look at recently proposed non-linear conjugate gradient algorithm with an efficient line search technique \citep{PalRoy21,PalRoy22}. An estimate of the covariance matrix of the ML estimate of $\hat {\bm \theta}$ can be obtained by
\begin{equation}\label{varcovm}
 \widehat{\text{Var}}(\hat{\bm \theta}) \approx -\left\{\frac{\partial^2 l(\bm \theta| \mathcal{D}_O)}{\partial \bm \theta \partial \bm \theta^{\intercal}}\right\}^{-1} \Bigg \vert_{\bm \theta=\hat{\bm \theta}}
\end{equation}
where $l(.|\mathcal{D}_O)$ is given in eqn. (\ref{lobs}). Standard errors (SEs) of the ML estimates are obtained by taking square-root of the diagonal elements of $\widehat{\text{Var}}(\hat{\bm \theta})$.

\section{Simulation study} \label{sec:sec4}

To illustrate the performance (robustness, accuracy, consistency and asymptotic normality) of the proposed model and estimation method under interval censored data, a detailed study is carried out under different parameter settings based on simulated data. The effects on the performance of our method due to variation in the parameter settings are captured by simulating data with transformation parameter $\alpha=0$ (PCM) , $0.5$ (BCTM) and $1$ (MCM), and $n=200$ (smaller sample size) and $400$ (moderate sample size). For $i=1, \dots, n$, $\bm x_i=(x_{1i}, x_{2i})^{\intercal}$ is the observed covariate vector for the $i$th subject where $x_{1i}$ is generated from a Bernoulli distribution with success probability $0.5$, and $x_{2i}$ is generated from a uniform distribution in the interval $[0.1, 20.0]$. Further, $\bm z_i=\bm x_i$ for $i=1, \dots, n$, and the true values of $\beta_0=0.6$, $\beta_1=-1.5$, $\beta_2=0.1$, $\gamma_1=-1.2$ and $\gamma_2=0.1$ are chosen conveniently. Assuming the true baseline hazard function to be a real constant ($\zeta>0$), corresponding to which the simulated data is generated, we obtain the common cumulative hazard function associated with the latency as $\Lambda(y_i; \bm x_i, \bm \gamma)= \zeta y_i e^{\gamma_1x_{1i}+\gamma_2x_{2i}}$ for some $y_i > 0$. Since, $S_u(.; \bm x_i, \bm z_i, \bm \theta^*)$ in eqn. (\ref{SusS}) is a proper survival function, it takes values in $[0, 1]$ and is uniformly distributed. Therefore, generating $u_i$ from a uniform distribution $[0, 1]$, we may obtain $y_i$ by solving eqn. (\ref{SusS}) for true value of $\bm \theta^*=(\alpha,\zeta, \beta_0, \beta_1, \beta_2, \gamma_1, \gamma_2)^{\intercal}$. Therefore, the time-to-events $y_i$s associated with latency are generated from an exponential distribution with rate $\zeta e^{\gamma_1x_{1i}+\gamma_2x_{2i}}$. Further, note that $\bm \theta=(\alpha, \psi_0, \dots, \psi_B, \beta_0, \beta_1, \beta_2, \gamma_1, \gamma_2)^{\intercal}$ is of order $(B+7)$ corresponding to the fitted model. Censoring times $c_i, i=1, \dots, n$, are generated from an exponential distribution with rate parameter $\zeta^*>0$ and $t_i=\min\{y_i, c_i\}$ is set. If $u_i \le \pi(\bm \theta; \bm z_i)$, then $l_i=c_i, r_i=\infty$ and $\delta_i=0$. If $u_i > \pi(\bm \theta; \bm z_i)$ and $t_i=c_i$, then $l_i=c_i, r_i=\infty$ and $\delta_i=0$. Finally, if $u_i > \pi(\bm \theta; \bm z_i)$ and $t_i=y_i$, then $\delta_i=1$. In this case, $d_1$ and $d_2$ are generated from uniform distribution in $[0.2, 0.7]$ and $[0,1]$, respectively. If $t_i \in (0, d_2)$, we set $l_i=0$ and $r_i=d_2$. If $t_i \ge d_2$, we set $l_i=d_2+(\rho-1)d_1$ and $r_i=d_2+\rho d_1$, where $\rho=\left \lfloor{\frac{t_i-d_2}{d_1}}\right \rfloor+1$ (\citealp{Pal17c}). $\zeta$ is chosen as $0.1$ or $0.2$ with $\zeta^*=0.1$ to maintain censoring rate of $20\%-30\%$ for every simulated scenario. 

The performance of our method is assessed in terms of measures related to estimation accuracy and consistency. These include parameter estimate (EST), standard error (SE), bias in estimation (BIAS), root mean squared error (RMSE), coverage probability (CP) at a conventional nominal level of 95\%. Further, model fit, and hence, performance of the BCTM model by incorporating PLA is assessed in terms of maximized log-likelihood value ($\hat l$) and Akaike information criterion (AIC). The above-mentioned measures are calculated based on 400 Monte-Carlo simulation runs. An initial guess for $\alpha$ is taken as $0.5$ since the search region for $\alpha$ is $[0, 1]$. The initial guesses for $\beta_0, \beta_1$, $\beta_2$, $\gamma_1$ and $\gamma_2$ are generated randomly from an interval with $10\%$ margin of error from their true values. For the simulation study, the cut-points or knots, $\tau_0, \dots,  \tau_B$, are considered as the equi-proportion quantiles of the finite lower and upper limits of the intervals. That is, $\tau_0=0$, $\tau_{B}=\max \{S_B\}$ where $S_B=\{l_i, r_i: l_i < \infty, r_i < \infty,  i=1, \dots,n \}$. For any $b \in \{1, \dots, B-1\}$, $\tau_b$ is taken as the  $\lfloor{ \frac{b}{B} }\rfloor$th quantile of $S_B$. First, all $\psi_b, b=0, \dots, B$, are taken as $0.001$. Then, for $b=1, \dots, B$, $\psi_b$ is set to $\psi_b \leftarrow: \psi_{b-1}+\psi_{b}+\psi_{b}^2$. These values provide initial estimates of $\psi_0, \dots, \psi_B$. The choices are considered keeping in mind that the hazard values may gradually and non-linearly increase with time at every cut-point.

For all six scenarios described above, i.e, $(\alpha, n)=$
(0.0, 200), (0.0, 400), (0.5, 200), (0.5, 400), (1.0, 200), (1.0, 400), for which data are simulated, the BCTM models with PLA-based baseline hazard function with $B=1, 2, 3$ and $4$ are fitted. The simulation study results are presented in Tables \ref{tab_sim1}-\ref{tab_sim6}. The Monte-Carlo simulation run based estimates are quite accurate with relatively similar standard error values (relative with respect to similar studies on cure models) for most parameters. The standard errors associated with $\beta_0$ are high compared to other parameters, most likely because $\beta_0$ is an intercept parameter which is susceptible to small fluctuations in data. The baseline hazard function is estimated by the PLA function in the time interval $[\tau_0, \tau_B]$ where often the Euclidean distance between $\tau_{B-1}$ and $\tau_B$ is large. Therefore, we may notice that standard error associated with $\psi_B$ is almost $3-6$ folds high when compared with respect to other $\psi_b, b=0, \dots, B-1$, owing to lesser accuracy of the PLA estimation at the last cut-point.  Further, the simulated data is generated from an exponential baseline distribution corresponding to the latency part of the model. Thus, the simulated data is generated with true time-to-event values in the interval $[0, \infty)$ already influenced by our interval censoring schemes define earlier, and the baseline hazard function is estimated by the PLA function in the partial time interval $[\tau_0, \tau_B]$ where $||\tau_{B}-\tau_{B-1}||_{_2}$ is large. These are some of the factors likely to be resulting relatively large value of SE corresponding to $\psi_B$. This phenomenon is along the lines of previous findings by \cite{Bala-Barui22}. The SE corresponding to $\alpha$ is high,  noting that it can only take values in $[0, 1]$. Moreover, in some settings, especially for true $\alpha=0.5$, the CPs show under-coverage (70.1863\%) even when average of the MLEs of $\alpha$ show closeness to 0.5 and SE is as high as $0.7806$. These can be attributed to the flatness of the likelihood function surface with respect to $\alpha$ resulting in fluctuating and inconsistent ML estimate for $\alpha$ with high variability. However, CPs corresponding to $\alpha$ are closer to the nominal level when data are generated from the PCTM ($\alpha=0.0$) or the MCM ($\alpha=1.0$). The SE for $\alpha$ is comparatively smaller when data are generated from $\alpha=0.0$ and higher when data is generated from $\alpha=1.0$. Even though for cases under interval-censored data with true $\alpha=0.5$ the CPs show under-coverage, we would like to emphasize that the BIAS in the estimation of $\alpha$ are not high, the ML estimates related to other parameters show satisfactory results (with respect to all measures considered) and the running time for the entire Monte-Carlo simulation runs (with 400 replications) is much less in comparison to the profile-likelihood approach to estimate $\alpha$. For profile-likelihood based technique, even for a basic grid search with $\alpha= 0.0, 0.1, \dots, 1.0$, it can take $5-10$ times greater computation time (running time) than the simultaneous-maximization based approach depending on the values of $\alpha$ and $n$. For better perspective, for interval-censored data generated from true $\alpha=0.5$ and $n=200$, the computation time related to estimation and model fitting is approximately 11.57 minutes by applying the profile-likelihood method with $\alpha \in  \{0.0, 0.1, \dots, 1.0\}$ while that of the simultaneous-maximization method is 1.49 minutes for a single Monte-Carlo run.

Some obvious trends are observed, e.g., increasing $n$ from $200$ to $400$ results in the overall decrease in SE, BIAS and RMSE,  while the CPs get closer to the nominal level for almost all settings. Finally, since our proposed BCTM with $B=b$ is nested in the BCTM with $B=b'$ for $b<b'$, the maximized log-likelihood value increases as $b$ increases. As expected, this trend has been captured in all simulation settings. Due to computational challenges, we did not continue with PLAs for $b \ge 5$.

\begin{table}[!htbp]
\caption{Estimation results based on simultaneous-maximization for interval censored data with $\alpha=0$ and $n=200$}\label{tab_sim1}
\resizebox{\columnwidth}{!}{
\begin{tabular}{c|cccccc|cccccc}
\hline
\multicolumn{1}{c|}{Fitted Model} &  Measure          &      $\psi_0$ &      $\psi_1$ &      $\psi_2$ &      $\psi_3$ &      $\psi_4$ &     $\beta_0 (0.6)$ &     $\beta_1 (-1.5)$ &     $\beta_2 (0.1)$ &    $\gamma_1 (-1.2) $ &    $\gamma_2 (0.1)$ &      $\alpha (0.0)$ \\
\hline
           &        EST &     0.5172 &     1.5040 & - & - & - &     0.6869 &    -1.6637 &     0.1180 &    -1.2918 &     0.0935 &     0.0730 \\

       $B=1$ &         SE &     0.2813 &     1.1945 & - & - & - &     0.4900 &     0.6763 &     0.0585 &     0.6499 &     0.0765 &     0.1748 \\

($\hat l$=-349.2401) &       BIAS & - & - & - & - & - &     0.0869 &    -0.1637 &     0.0180 &    -0.0918 &    -0.0065 &     0.0730 \\

           &       RMSE & - & - & - & - & - &     0.4976 &     0.6958 &     0.0612 &     0.6564 &     0.0768 &     0.1894 \\

           &  CP (95\%) & - & - & - & - & - &    97.9798 &   100.0000 &    99.4949 &    93.4343 &    94.9495 &    98.9899 \\
\hline
           &        EST &     0.6205 &     0.6226 &     1.7510 & - & - &     0.7067 &    -1.7844 &     0.1215 &    -1.1699 &     0.0923 &     0.0929 \\

       $B=2$ &         SE &     0.4930 &     0.4143 &     1.8544 & - & - &     0.6074 &     0.7221 &     0.0705 &     0.7260 &     0.0786 &     0.2714 \\

($\hat l$=-348.4369) &       BIAS & - & - & - & - & - &     0.1067 &    -0.2844 &     0.0215 &     0.0301 &    -0.0077 &     0.0929 \\

           &       RMSE & - & - & - & - & - &     0.6167 &     0.7761 &     0.0737 &     0.7266 &     0.0790 &     0.2869 \\

           &  CP (95\%) & - & - & - & - & - &    97.4747 &    99.4949 &    98.9899 &    95.4545 &    94.9495 &    98.9899 \\
\hline
           &        EST &     0.5840 &     0.6874 &     0.6372 &     1.8925 & - &     0.6959 &    -1.7874 &     0.1220 &    -1.1568 &     0.0906 &     0.0879 \\

       $B=3$ &         SE &     0.4221 &     0.4058 &     0.3773 &     1.6333 & - &     0.4783 &     0.5749 &     0.0602 &     0.5586 &     0.0711 &     0.1947 \\

($\hat l$=-348.1014) &       BIAS & - & - & - & - & - &     0.0959 &    -0.2874 &     0.0220 &     0.0432 &    -0.0094 &     0.0879 \\

           &       RMSE & - & - & - & - & - &     0.4878 &     0.6427 &     0.0641 &     0.5603 &     0.0717 &     0.2136 \\

           &  CP (95\%) & - & - & - & - & - &    95.9596 &    98.4848 &    98.9899 &    88.3838 &    96.9697 &   100.0000 \\
\hline
           &        EST &     0.6737 &     0.6275 &     0.8246 &     0.7059 &     2.2006 &     0.6556 &    -1.7610 &     0.1211 &    -1.1348 &     0.0847 &     0.0772 \\

       $B=4$ &         SE &     0.5331 &     0.4265 &     0.5459 &     0.4641 &     2.1660 &     0.4546 &     0.5728 &     0.0561 &     0.5750 &     0.0734 &     0.1988 \\

($\hat l$=-347.9270) &       BIAS & - & - & - & - & - &     0.0556 &    -0.2610 &     0.0211 &     0.0652 &    -0.0153 &     0.0772 \\

           &       RMSE & - & - & - & - & - &     0.4580 &     0.6295 &     0.0599 &     0.5787 &     0.0750 &     0.2133 \\

           &  CP (95\%) & - & - & - & - & - &    93.9394 &    97.9798 &    99.4949 &    90.4040 &    96.9697 &    99.4949 \\
\hline
\end{tabular}} \\
\footnotesize{$\hat l$:maximized log-likelihood value; true parameter value is given within parenthesis beside every parameter notation}
\end{table}

\begin{table}[!htbp]
\caption{Estimation results based on simultaneous-maximization for interval censored data with $\alpha=0$ and $n=400$}\label{tab_sim2}
\resizebox{\columnwidth}{!}{
\begin{tabular}{c|cccccc|cccccc}
\hline
\multicolumn{1}{c|}{Fitted Model} &  Measure          &      $\psi_0$ &      $\psi_1$ &      $\psi_2$ &      $\psi_3$ &      $\psi_4$ &     $\beta_0 (0.6)$ &     $\beta_1 (-1.5)$ &     $\beta_2 (0.1)$ &    $\gamma_1 (-1.2) $ &    $\gamma_2 (0.1)$ &      $\alpha (0.0)$ \\
\hline
           &        EST &     0.5054 &     1.4701 & - & - & - &     0.6327 &    -1.5834 &     0.1115 &    -1.2821 &     0.0914 &     0.0400 \\

       $B=1$ &         SE &     0.1821 &     0.7595 & - & - & - &     0.2915 &     0.3596 &     0.0410 &     0.4424 &     0.0631 &     0.1226 \\

($\hat l$=-624.2370) &       BIAS & - & - & - & - & - &     0.0327 &    -0.0834 &     0.0115 &    -0.0821 &    -0.0086 &     0.0400 \\

           &       RMSE & - & - & - & - & - &     0.2933 &     0.3691 &     0.0426 &     0.4500 &     0.0637 &     0.1290 \\

           &  CP (95\%) & - & - & - & - & - &    97.0213 &    98.7234 &    98.7234 &    95.7447 &    98.2979 &    99.1489 \\
\hline
           &        EST &     0.5995 &     0.6091 &     1.6515 & - & - &     0.6306 &    -1.6514 &     0.1135 &    -1.1654 &     0.0894 &     0.0477 \\

       $B=2$ &         SE &     0.3000 &     0.2739 &     1.2268 & - & - &     0.3716 &     0.4342 &     0.0451 &     0.5349 &     0.0677 &     0.1591 \\

($\hat l$=-623.2617) &       BIAS & - & - & - & - & - &     0.0306 &    -0.1514 &     0.0135 &     0.0346 &    -0.0106 &     0.0477 \\

           &       RMSE & - & - & - & - & - &     0.3729 &     0.4598 &     0.0471 &     0.5360 &     0.0685 &     0.1661 \\

           &  CP (95\%) & - & - & - & - & - &    94.8936 &    98.7234 &    98.2979 &    96.1702 &    96.5957 &    99.1489 \\
\hline
           &        EST &     0.5574 &     0.6798 &     0.6073 &     1.7950 & - &     0.5964 &    -1.6519 &     0.1142 &    -1.1272 &     0.0866 &     0.0390 \\

       $B=3$ &         SE &     0.3280 &     0.3221 &     0.2984 &     1.3334 & - &     0.3566 &     0.4091 &     0.0426 &     0.5048 &     0.0604 &     0.1517 \\

($\hat l$=-623.0527) &       BIAS & - & - & - & - & - &    -0.0036 &    -0.1519 &     0.0142 &     0.0728 &    -0.0134 &     0.0390 \\

           &       RMSE & - & - & - & - & - &     0.3566 &     0.4364 &     0.0449 &     0.5100 &     0.0619 &     0.1566 \\

           &  CP (95\%) & - & - & - & - & - &    95.3191 &    99.1489 &    98.7234 &    97.0213 &    94.8936 &    98.7234 \\
\hline
           &        EST &     0.6382 &     0.6337 &     0.7832 &     0.6685 &     2.1034 &     0.5744 &    -1.6703 &     0.1175 &    -1.1202 &     0.0821 &     0.0451 \\

       $B=4$ &         SE &     0.3935 &     0.3109 &     0.4065 &     0.3316 &     1.4735 &     0.3358 &     0.4057 &     0.0404 &     0.4418 &     0.0604 &     0.1595 \\

($\hat l$=-622.8717) &       BIAS & - & - & - & - & - &    -0.0256 &    -0.1703 &     0.0175 &     0.0798 &    -0.0179 &     0.0451 \\

           &       RMSE & - & - & - & - & - &     0.3368 &     0.4400 &     0.0440 &     0.4489 &     0.0630 &     0.1658 \\

           &  CP (95\%) & - & - & - & - & - &    92.7660 &    98.2979 &    97.0213 &    92.7660 &    93.6170 &    99.1489 \\

\hline
\end{tabular}} \\
\footnotesize{$\hat l$:maximized log-likelihood value; true parameter value is given within parenthesis beside every parameter notation}
\end{table}

\begin{table}[!htbp]
\caption{Estimation results based on simultaneous-maximization for interval censored data with $\alpha=0.5$ and $n=200$}\label{tab_sim3}
\resizebox{\columnwidth}{!}{
\begin{tabular}{c|cccccc|cccccc}
\hline
\multicolumn{1}{c|}{Fitted Model} &  Measure          &      $\psi_0$ &      $\psi_1$ &      $\psi_2$ &      $\psi_3$ &      $\psi_4$ &     $\beta_0 (0.6)$ &     $\beta_1 (-1.5)$ &     $\beta_2 (0.1)$ &    $\gamma_1 (-1.2) $ &    $\gamma_2 (0.1)$ &      $\alpha (0.5)$ \\
\hline
           &        EST &     0.4270 &     1.0540 & - & - & - &     0.8800 &    -1.2840 &     0.0972 &    -1.0500 &     0.0829 &     0.3564 \\

       $B=1$ &         SE &     0.2337 &     0.5614 & - & - & - &     0.7073 &     0.6828 &     0.0607 &     0.4187 &     0.0485 &     0.4247 \\

($\hat l$=-445.0887) &       BIAS & - & - & - & - & - &     0.2800 &     0.2160 &    -0.0028 &     0.1500 &    -0.0171 &    -0.1436 \\

           &       RMSE & - & - & - & - & - &     0.7607 &     0.7162 &     0.0608 &     0.4448 &     0.0514 &     0.4483 \\

           &  CP (95\%) & - & - & - & - & - &    90.6832 &    88.1988 &    96.2733 &    93.7888 &    94.4099 &    58.3851 \\
\hline
           &        EST &     0.5229 &     0.4884 &     1.0533 & - & - &     1.1078 &    -1.5495 &     0.1016 &    -1.0756 &     0.0884 &     0.4741 \\

       $B=2$ &         SE &     0.3857 &     0.2549 &     0.6939 & - & - &     1.0130 &     0.9813 &     0.0767 &     0.4354 &     0.0440 &     0.6402 \\

($\hat l$=-443.9937) &       BIAS & - & - & - & - & - &     0.5078 &    -0.0495 &     0.0016 &     0.1244 &    -0.0116 &    -0.0259 \\

           &       RMSE & - & - & - & - & - &     1.1332 &     0.9825 &     0.0767 &     0.4528 &     0.0455 &     0.6407 \\

           &  CP (95\%) & - & - & - & - & - &    95.0311 &    91.3043 &    97.5155 &    93.1677 &    95.6522 &    69.5652 \\
\hline
           &        EST &     0.4826 &     0.5521 &     0.4784 &     1.1552 & - &     1.0696 &    -1.5738 &     0.1032 &    -1.0525 &     0.0864 &     0.4678 \\

       $B=3$ &         SE &     0.3969 &     0.3656 &     0.2619 &     0.8104 & - &     1.0413 &     1.0072 &     0.0751 &     0.4549 &     0.0458 &     0.6714 \\

($\hat l$=-443.4984) &       BIAS & - & - & - & - & - &     0.4696 &    -0.0738 &     0.0032 &     0.1475 &    -0.0136 &    -0.0322 \\

           &       RMSE & - & - & - & - & - &     1.1423 &     1.0099 &     0.0752 &     0.4782 &     0.0478 &     0.6722 \\

           &  CP (95\%) & - & - & - & - & - &    95.0311 &    91.9255 &    96.2733 &    94.4099 &    96.2733 &    73.2919 \\
\hline
           &        EST &     0.5475 &     0.5231 &     0.6056 &     0.4999 &     1.4598 &     1.0413 &    -1.5915 &     0.1088 &    -1.0767 &     0.0860 &     0.5040 \\

       $B=4$ &         SE &     0.5380 &     0.4108 &     0.4184 &     0.3228 &     1.1112 &     1.1934 &     1.1011 &     0.0834 &     0.5023 &     0.0544 &     0.7806 \\

($\hat l$=-443.1375) &       BIAS & - & - & - & - & - &     0.4413 &    -0.0915 &     0.0088 &     0.1233 &    -0.0140 &     0.0040 \\

           &       RMSE & - & - & - & - & - &     1.2724 &     1.1049 &     0.0839 &     0.5172 &     0.0562 &     0.7806 \\

           &  CP (95\%) & - & - & - & - & - &    95.0311 &    92.5466 &    98.1366 &    93.7888 &    96.8944 &    70.1863 \\

\hline
\end{tabular}} \\
\footnotesize{$\hat l$:maximized log-likelihood value; true parameter value is given within parenthesis beside every parameter notation}
\end{table}

\begin{table}[!htbp]
\caption{Estimation results based on simultaneous-maximization for interval censored data with $\alpha=0.5$ and $n=400$}\label{tab_sim4}
\resizebox{\columnwidth}{!}{
\begin{tabular}{c|cccccc|cccccc}
\hline
\multicolumn{1}{c|}{Fitted Model} &  Measure          &      $\psi_0$ &      $\psi_1$ &      $\psi_2$ &      $\psi_3$ &      $\psi_4$ &     $\beta_0 (0.6)$ &     $\beta_1 (-1.5)$ &     $\beta_2 (0.1)$ &    $\gamma_1 (-1.2) $ &    $\gamma_2 (0.1)$ &      $\alpha (0.5)$ \\
\hline
           &        EST &     0.4250 &     1.0440 & - & - & - &     0.9013 &    -1.3100 &     0.0910 &    -1.1037 &     0.0851 &     0.3670 \\

       $B=1$ &         SE &     0.1912 &     0.4544 & - & - & - &     0.6132 &     0.5612 &     0.0472 &     0.3149 &     0.0376 &     0.4036 \\

($\hat l$=-806.8103) &       BIAS & - & - & - & - & - &     0.3013 &     0.1900 &    -0.0090 &     0.0963 &    -0.0149 &    -0.1330 \\

           &       RMSE & - & - & - & - & - &     0.6832 &     0.5925 &     0.0481 &     0.3293 &     0.0404 &     0.4249 \\

           &  CP (95\%) & - & - & - & - & - &    89.3939 &    81.3131 &    93.9394 &    93.4343 &    90.9091 &    64.6465 \\
\hline
           &        EST &     0.4869 &     0.4656 &     0.9415 & - & - &     1.0040 &    -1.4111 &     0.0960 &    -1.1053 &     0.0923 &     0.4447 \\

       $B=2$ &         SE &     0.3571 &     0.2388 &     0.5423 & - & - &     0.9001 &     0.8594 &     0.0666 &     0.3610 &     0.0367 &     0.6631 \\

($\hat l$=-805.0873) &       BIAS & - & - & - & - & - &     0.4040 &     0.0889 &    -0.0040 &     0.0947 &    -0.0077 &    -0.0553 \\

           &       RMSE & - & - & - & - & - &     0.9866 &     0.8640 &     0.0667 &     0.3732 &     0.0375 &     0.6654 \\

           &  CP (95\%) & - & - & - & - & - &    95.4545 &    88.8889 &    95.4545 &    92.4242 &    93.4343 &    72.2222 \\
\hline
           &        EST &     0.4456 &     0.5339 &     0.4546 &     1.0131 & - &     1.0273 &    -1.4180 &     0.0971 &    -1.1230 &     0.0929 &     0.4637 \\

       $B=3$ &         SE &     0.2952 &     0.2756 &     0.2032 &     0.5578 & - &     0.7732 &     0.7431 &     0.0622 &     0.3216 &     0.0344 &     0.5513 \\

($\hat l$=-804.4456) &       BIAS & - & - & - & - & - &     0.4273 &     0.0820 &    -0.0029 &     0.0770 &    -0.0071 &    -0.0363 \\

           &       RMSE & - & - & - & - & - &     0.8834 &     0.7476 &     0.0623 &     0.3307 &     0.0351 &     0.5525 \\

           &  CP (95\%) & - & - & - & - & - &    93.9394 &    87.3737 &    93.4343 &    92.9293 &    92.4242 &    72.2222 \\
\hline
           &        EST &     0.5106 &     0.5287 &     0.5697 &     0.5005 &     1.2440 &     1.0356 &    -1.5389 &     0.1071 &    -1.1143 &     0.0903 &     0.5244 \\

       $B=4$ &         SE &     0.3094 &     0.2780 &     0.2576 &     0.2102 &     0.6712 &     0.7485 &     0.7416 &     0.0659 &     0.3075 &     0.0337 &     0.5325 \\

($\hat l$=-804.2285) &       BIAS & - & - & - & - & - &     0.4356 &    -0.0389 &     0.0071 &     0.0857 &    -0.0097 &     0.0244 \\

           &       RMSE & - & - & - & - & - &     0.8660 &     0.7426 &     0.0663 &     0.3192 &     0.0351 &     0.5331 \\

           &  CP (95\%) & - & - & - & - & - &    92.4242 &    87.8788 &    92.4242 &    92.9293 &    94.9495 &    71.7172 \\

\hline
\end{tabular}} \\
\footnotesize{$\hat l$:maximized log-likelihood value; true parameter value is given within parenthesis beside every parameter notation}
\end{table}

\begin{table}[!htbp]
\caption{Estimation results based on simultaneous-maximization for interval censored data with $\alpha=1.0$ and $n=200$}\label{tab_sim5}
\resizebox{\columnwidth}{!}{
\begin{tabular}{c|cccccc|cccccc}
\hline
\multicolumn{1}{c|}{Fitted Model} &  Measure          &      $\psi_0$ &      $\psi_1$ &      $\psi_2$ &      $\psi_3$ &      $\psi_4$ &     $\beta_0 (0.6)$ &     $\beta_1 (-1.5)$ &     $\beta_2 (0.1)$ &    $\gamma_1 (-1.2) $ &    $\gamma_2 (0.1)$ &      $\alpha (1.0)$ \\
\hline
           &        EST &     0.3948 &     0.9057 & - & - & - &     1.1789 &    -1.1958 &     0.0896 &    -1.1161 &     0.0911 &     0.7008 \\

       $B=1$ &         SE &     0.3000 &     0.4528 & - & - & - &     1.2021 &     1.0935 &     0.0786 &     0.3770 &     0.0360 &     0.8461 \\

($\hat l$=-461.8779) &       BIAS & - & - & - & - & - &     0.5789 &     0.3042 &    -0.0104 &     0.0839 &    -0.0089 &    -0.2992 \\

           &       RMSE & - & - & - & - & - &     1.3342 &     1.1350 &     0.0793 &     0.3862 &     0.0371 &     0.8974 \\

           &  CP (95\%) & - & - & - & - & - &    91.7355 &    87.6033 &    94.2149 &    85.9504 &    94.2149 &    93.3884 \\
\hline
           &        EST &     0.4753 &     0.4471 &     1.0615 & - & - &     1.1364 &    -1.3398 &     0.0979 &    -1.0677 &     0.0880 &     0.7191 \\

       $B=2$ &         SE &     0.3493 &     0.2433 &     0.5766 & - & - &     1.0127 &     0.9163 &     0.0801 &     0.3462 &     0.0381 &     0.8035 \\

($\hat l$=-461.3796) &       BIAS & - & - & - & - & - &     0.5364 &     0.1602 &    -0.0021 &     0.1323 &    -0.0120 &    -0.2809 \\

           &       RMSE & - & - & - & - & - &     1.1460 &     0.9302 &     0.0801 &     0.3706 &     0.0399 &     0.8512 \\

           &  CP (95\%) & - & - & - & - & - &    90.0826 &    90.9091 &    90.9091 &    81.8182 &    97.5207 &    93.3884 \\
\hline
           &        EST &     0.4435 &     0.5543 &     0.4523 &     1.2320 & - &     1.1397 &    -1.3759 &     0.1028 &    -1.0781 &     0.0868 &     0.7701 \\

       $B=3$ &         SE &     0.3485 &     0.3513 &     0.2428 &     0.6921 & - &     1.1781 &     1.0035 &     0.0838 &     0.3249 &     0.0364 &     0.9184 \\

($\hat l$=-461.0614) &       BIAS & - & - & - & - & - &     0.5397 &     0.1241 &     0.0028 &     0.1219 &    -0.0132 &    -0.2299 \\

           &       RMSE & - & - & - & - & - &     1.2958 &     1.0111 &     0.0838 &     0.3470 &     0.0387 &     0.9467 \\

           &  CP (95\%) & - & - & - & - & - &    93.3884 &    90.0826 &    93.3884 &    85.1240 &    95.8678 &    93.3884 \\
\hline
           &        EST &     0.4911 &     0.5453 &     0.5815 &     0.5069 &     1.5417 &     1.1136 &    -1.4256 &     0.1079 &    -1.0646 &     0.0843 &     0.8036 \\

       $B=4$ &         SE &     0.4779 &     0.4742 &     0.4246 &     0.3336 &     1.1267 &     1.8993 &     1.5842 &     0.0920 &     0.3959 &     0.0394 &     1.4515 \\

($\hat l$=-460.8642) &       BIAS & - & - & - & - & - &     0.5136 &     0.0744 &     0.0079 &     0.1354 &    -0.0157 &    -0.1964 \\

           &       RMSE & - & - & - & - & - &     1.9675 &     1.5859 &     0.0923 &     0.4184 &     0.0424 &     1.4647 \\

           &  CP (95\%) & - & - & - & - & - &    90.0826 &    90.0826 &    95.0413 &    83.4711 &    92.5620 &    90.9091 \\

\hline
\end{tabular}} \\
\footnotesize{$\hat l$:maximized log-likelihood value; true parameter value is given within parenthesis beside every parameter notation}
\end{table}

\begin{table}[!htbp]
\caption{Estimation results based on simultaneous-maximization for interval censored data with $\alpha=1.0$ and $n=400$}\label{tab_sim6}
\resizebox{\columnwidth}{!}{
\begin{tabular}{c|cccccc|cccccc}
\hline
\multicolumn{1}{c|}{Fitted Model} &  Measure          &      $\psi_0$ &      $\psi_1$ &      $\psi_2$ &      $\psi_3$ &      $\psi_4$ &     $\beta_0 (0.6)$ &     $\beta_1 (-1.5)$ &     $\beta_2 (0.1)$ &    $\gamma_1 (-1.2) $ &    $\gamma_2 (0.1)$ &      $\alpha (1.0)$ \\
\hline
           &        EST &     0.3929 &     0.8482 & - & - & - &     1.1888 &    -1.1847 &     0.0777 &    -1.1340 &     0.0923 &     0.6572 \\

       $B=1$ &         SE &     0.2168 &     0.3106 & - & - & - &     0.9049 &     0.6918 &     0.0603 &     0.2627 &     0.0311 &     0.6714 \\

($\hat l$=-840.1980) &       BIAS & - & - & - & - & - &     0.5888 &     0.3153 &    -0.0223 &     0.0660 &    -0.0077 &    -0.3428 \\

           &       RMSE & - & - & - & - & - &     1.0796 &     0.7603 &     0.0643 &     0.2709 &     0.0320 &     0.7538 \\

           &  CP (95\%) & - & - & - & - & - &    94.6565 &    75.5725 &    84.7328 &    78.6260 &    96.1832 &    87.7863 \\
\hline
           &        EST &     0.4592 &     0.4319 &     0.9438 & - & - &     1.1516 &    -1.2637 &     0.0850 &    -1.1072 &     0.0913 &     0.6875 \\

       $B=2$ &         SE &     0.2799 &     0.1938 &     0.3899 & - & - &     0.8522 &     0.6860 &     0.0635 &     0.2638 &     0.0286 &     0.6964 \\

($\hat l$=-839.4499) &       BIAS & - & - & - & - & - &     0.5516 &     0.2363 &    -0.0150 &     0.0928 &    -0.0087 &    -0.3125 \\

           &       RMSE & - & - & - & - & - &     1.0151 &     0.7256 &     0.0652 &     0.2796 &     0.0299 &     0.7633 \\

           &  CP (95\%) & - & - & - & - & - &    93.8931 &    81.6794 &    84.7328 &    81.6794 &    93.8931 &    91.6031 \\
\hline
           &        EST &     0.4413 &     0.5316 &     0.4431 &     1.0931 & - &     1.1449 &    -1.3442 &     0.0925 &    -1.0952 &     0.0878 &     0.7402 \\

       $B=3$ &         SE &     0.3033 &     0.2868 &     0.2078 &     0.4758 & - &     0.9398 &     0.7881 &     0.0714 &     0.2688 &     0.0314 &     0.8145 \\

($\hat l$=-839.4347) &       BIAS & - & - & - & - & - &     0.5449 &     0.1558 &    -0.0075 &     0.1048 &    -0.0122 &    -0.2598 \\

           &       RMSE & - & - & - & - & - &     1.0863 &     0.8034 &     0.0718 &     0.2885 &     0.0337 &     0.8549 \\

           &  CP (95\%) & - & - & - & - & - &    91.6031 &    77.0992 &    84.7328 &    83.9695 &    93.1298 &    84.7328 \\
\hline
           &        EST &     0.5081 &     0.5440 &     0.5780 &     0.4949 &     1.4524 &     1.1416 &    -1.4273 &     0.1003 &    -1.0866 &     0.0843 &     0.8086 \\

       $B=4$ &         SE &     0.2909 &     0.2372 &     0.2263 &     0.1770 &     0.5884 &     0.7951 &     0.6379 &     0.0689 &     0.2126 &     0.0246 &     0.6582 \\

($\hat l$=-839.4547) &       BIAS & - & - & - & - & - &     0.5416 &     0.0727 &     0.0003 &     0.1134 &    -0.0157 &    -0.1914 \\

           &       RMSE & - & - & - & - & - &     0.9620 &     0.6420 &     0.0689 &     0.2410 &     0.0292 &     0.6855 \\

           &  CP (95\%) & - & - & - & - & - &    91.6031 &    81.6794 &    90.0763 &    88.5496 &    88.5496 &    83.9695 \\

\hline
\end{tabular}} \\
\footnotesize{$\hat l$:maximized log-likelihood value; true parameter value is given within parenthesis beside every parameter notation}
\end{table}


\section{Analysis of real data from a smoking cessation study} \label{sec:sec5}

The performance of our proposed model and estimation method is further demonstrated through an application on interval censored data obtained from the well-known smoking cessation study (\citealp{murray1998effects}). The main objective of the study is to illustrate potency of an aggressive intervention therapy on patients with nicotine addiction. The analysis data set contains intervals (`Timept1' and `Timept2' denoting the left and right limits) in which actual times to relapse occur for 223 subjects with history of nicotine addiction. These subjects were randomly assigned to either the smoking intervention (SI) therapy group (denoted by `SI.UC'=1) or the usual care (UC) group (`SI.UC'=0). Available covariate information include duration of smoking in years prior to the study enrollment (`Duration') and average numbers of cigarettes smoked per day prior to the study enrollment (`F10Cigs'). If a subject experienced relapse then the variable `Relapse', present in the data set, takes the value 1, else `Relapse' takes the value 0. `Timept2' is `NA' (which means infinite time to relapse) for `Relapse'=0. A descriptive summary of the data with some relevant measures is provided in Table \ref{real_desc}.

\begin{table}[!htbp]
\caption{Smoking cessation study: Descriptive summary table}\label{real_desc}
\centering
\begin{tabular}{ccccc}
\hline
           & Treatment Group &     Female &       Male &     Total  \\
\hline
\multicolumn{ 1}{c}{Relapse (\%)} &    SI.UC=1 &      32.88 &      21.88 &      26.63 \\

\multicolumn{ 1}{c}{} &    SI.UC=0 &      35.71 &       37.50 &      37.04 \\
\hline
\multicolumn{ 1}{c}{Median Relapse} &    SI.UC=1 &      4.8790 &      4.9030 &     4.8915 \\

\multicolumn{ 1}{c}{Time (in years)} &    SI.UC=0 &      4.9250 &      4.8980 &     4.9075 \\
\hline
\multicolumn{ 1}{c}{*Avg. Duration (sd) } &    SI.UC=1 &    29.5068 (6.3904) &    30.3438 (7.1153) &    29.9822 (6.8046) \\

\multicolumn{ 1}{c}{} &    SI.UC=0 &    28.2143 (8.8333) &      30.7500 (7.5030) &    30.0926 (7.8627) \\
\hline
\multicolumn{ 1}{c}{**Avg. F10Cigs (sd)} &    SI.UC=1 &    25.2466 (9.6678) &     29.3750 (12.5524) &    27.5917 (11.5482) \\

\multicolumn{ 1}{c}{} &    SI.UC=0 &    22.7143 (9.1604) &     26.8750 (9.9155) &    25.7963 (9.8141) \\
\hline
\end{tabular} \\ 
\footnotesize{*Avg. Duration: Average duration (in years) of smoking prior to the study enrollment \\ **Avg. F10Cigs: Average number of cigarettes smoked per day prior to the study enrollment\\
sd: standard deviation}
\end{table}
In our analysis, `Duration' ($x_1$; continuous), `F10Cigs' ($x_2$; continuous) and `SI.UC' ($x_3$; categorical) are considered as covariates. The proposed BCTM model with $B=1, \dots, 6$ are fitted to the smoking cessation data and their model adequacies are compared with respect to the maximized log-likelihood value ($\hat l$) and Akaike information criterion (AIC) (see Table \ref{real_aic}). For user-defined $B$, the initial parameter values, all denoted by notations used earlier with subscript `0', are chosen in the following sequential manner:
\begin{itemize}
    \item [(i)] $\alpha_0=0.5$ chosen as the mid-point of the search interval $[0,1]$.
    \item [(ii)] Some logical assumptions are made to obtain $\bm \beta_0$. Subjects with higher values of $x_1$ and $x_2$ are less likely to be cured. Similarly, subjects with smaller values of $x_1$ and $x_2$ are more likely to be cured. The observed censoring proportion for the smoking intervention and usual care groups are $0.7337$ and $0.6296$, respectively. Therefore, a subject with minimum values of $x_1$ and $x_2$ is more likely to have cure probability estimate as $0.7337$ (if belongs to `SI.UC=1') or $0.6296$ (if belongs to `SI.UC=0'). Note that, censoring proportion is the maximum attainable value for cure probability estimate for a given data. Further, halved cure probability estimates, i.e., $0.3669$ (`SI.UC=1') or $0.3148$ (`SI.UC=0') can be assigned to a subject with average values of $x_1$ and $x_2$, respectively. Simultaneously solving the four equations mentioned above leads to $\bm \beta_0 = (0.5819,  0.3131,  0.6416, -1.1478)^{\intercal}$.
    \item [(iii)] For unique finite time points, $t_1, \dots, t_p$, let $\hat{S}_{1}, \dots, \hat{S}_{p}$ be the Kaplan-Meier estimates for interval censored data (\citealp{lindsey1998methods}). Using Euclidean distance measure and some approximations, we obtain $\bm x_j=(x_{1j},x_{2j},x_{3j})^{\intercal}$ for $j=1, \dots, p$ from the smoking cessation data. By our assumption of proportional hazards structure, we get $\hat{S}_{j}=\{S_{0j}\}^{\exp\{\bm x_j^{\intercal} \bm \gamma\}}$, and hence, the transformed equation is 
\begin{equation} \label{EQ1}
\ln\{-\ln \{\hat{S}_{j}\}\}={\exp\{\bm x_j^{\intercal} \bm \gamma\}}+\ln\{-\ln \{S_{0}\}\}
\end{equation}
where $S_{0}=S_{0{j}}$ for $j=1, \dots, p$ is the baseline survival function assumed constant over time. Applying multiple linear regression to the data $$\{({\hat S}_j, \bm x_j); j=1, \dots, p\},$$ we obtain $\bm \gamma_0=(-0.2227,-0.0042,0.0185)^{\intercal}$. 
\item [(iv)] From eqn. (\ref{EQ1}), we get ${\hat S}_{0j}$ (no longer assumed to be constant over time), and hence, ${\hat H}_j=-\ln\{{\hat S}_j\}$ for $j=1, \dots, p$ by eliminating the effect of $\bm x_j^{\intercal}\bm \gamma_0$. From the set of paired numbers $\{(t_j, {\hat H}_j); j=1, \dots, p\}$, we may obtain the set of paired numbers $\{(t_j, {\hat h}_j); j=1, \dots, (p-1)\}$ where ${\hat h}_j=\frac{{\hat H}_{j+1}-{\hat H}_{j}}{t_{j+1}-t_{j}}$. Here, ${\hat h}_j$ is an approximation to the baseline hazard function value at $t_j$.
\item [(v)] We may plot $\{(t_j, {\hat h}_j); j=1, \dots, (p-1)\}$ to decide on our choice of the cut-points $\{\tau_0, \dots, \tau_B\}$ and corresponding initial parameter estimate $\bm \psi_0$ for fixed $B$. $\tau_0=0$ and $\tau_B$ is taken as the maximum value of the observed finite left and right limits of the time intervals. The cut-points are decided based on the points of inflections of the plotted graph. Other strategy that one may take is by considering suitable quantiles of the observed interval limit values.      
\end{itemize}

It is observed that the maximized log-likelihood values increase with an increase in $B$ (see Table \ref{real_aic}). This model behavior is expected since the cut-points are chosen in a way that the resulting model becomes a part of the family of nested models, i.e., BCTM with $B=b$ is nested within BCTM with $B=b'$ for $b<b'$. AIC, on the other hand, does not show monotonic behavior necessarily. Findings for the BCTM models with $B>6$ are not presented here since increasing $B$ further does not show significant decrease in AIC. The model parameter estimates and corresponding standard errors are presented in Table \ref{real_est}. Note that $x_1$ and $x_2$ show negative relationship while $x_3$ shows a positive relationship with the cure rate (or, in this case, non-relapse rate). Most of the parameter and standard error estimates seem reasonable. However, a higher variability is observed to be associated with $\psi_5$, that is for the baseline hazard function value corresponding to the last cut-point. This point has already been discussed in details in Section \ref{sec:sec4}. The ML estimate of $\alpha$ is obtained as 1.0000 (SE = 0.0001). This indicates that for the smoking cessation interval censored data, the MCM with $B=6$ gives the best fit in terms of AIC. Further, it is observed that baseline hazards increase in the intermediate cut-points of the study, however, shows decreasing trend in the later cut-points, except, for the last cut-point which provides the highest hazard value.            

\begin{table}[!htbp]
\caption{Maximized log-likelihood ($\hat l$) and Akaike Information Criterion (AIC) values obtained with respect to different degree of non-parametricity of the candidate models fitted to the smoking cessation data}\label{real_aic}
\centering
\begin{tabular}{c c c c}
\hline
{Degree of non-parametricity } &  $\hat l$ &    {$p_n$} &  { AIC} \\
\hline
$B=1$ & -199.5193	& 10	& 419.0386\\
$B=2$ & -199.4452	& 11	& 420.8903\\
$B=3$ & -198.1549	& 12	& 420.3097\\
$B=4$ & -196.3150 	&13	   & 418.6300\\
$B=5$ & -195.2093	&14	& 418.4185\\
$B=6$ & -194.1805	& 15	&418.3610\\
\hline
\end{tabular}\\  
\footnotesize{$p_n$: Number of estimated parameters
}
\end{table}

\begin{table}[!htbp]
\caption{Estimation results obtained by fitting our proposed model with degree of non-parametricity $B=6$ to the smoking cessation interval censored data}\label{real_est}
\centering
\begin{tabular}{ccccccccc}
\hline
{Parameter} & {$\psi_0$} & {$\psi_1$} & {$\psi_2$} & {$\psi_3$} & {$\psi_4$} & {$\psi_5$} & {$\psi_6$} & {$\beta_0$} \\
\hline
       EST &     0.0005 &     0.3973 &     0.0003 &     0.2181 &     0.1793 &     0.0005 &     5.3528 &     0.2248 \\

        SE &     0.0061 &     0.1138 &     0.2216 &     0.0117 &     0.2848 &     0.0068 &     4.5512 &     0.2027 \\ 
\hline 
\\

\hline
{Parameter} & {$\beta_1$} & {$\beta_2$} & {$\beta_3$} & {$\gamma_1$} & { $\gamma_2$} & {$\gamma_3$} & {$\alpha$} &            \\
\hline
       EST &    -0.5731 &    -0.2999 &     0.9061 &    -0.9483 &     1.6274 &    -0.1742 &     1.0000 &            \\

        SE &     0.1058 &     0.4107 &     0.4758 &     0.4729 &     0.5332 &     0.7130 &     0.0001 &            \\
\hline
\end{tabular}  
\end{table}

\begin{figure}[!htbp]
\centering
\includegraphics[width=6in]{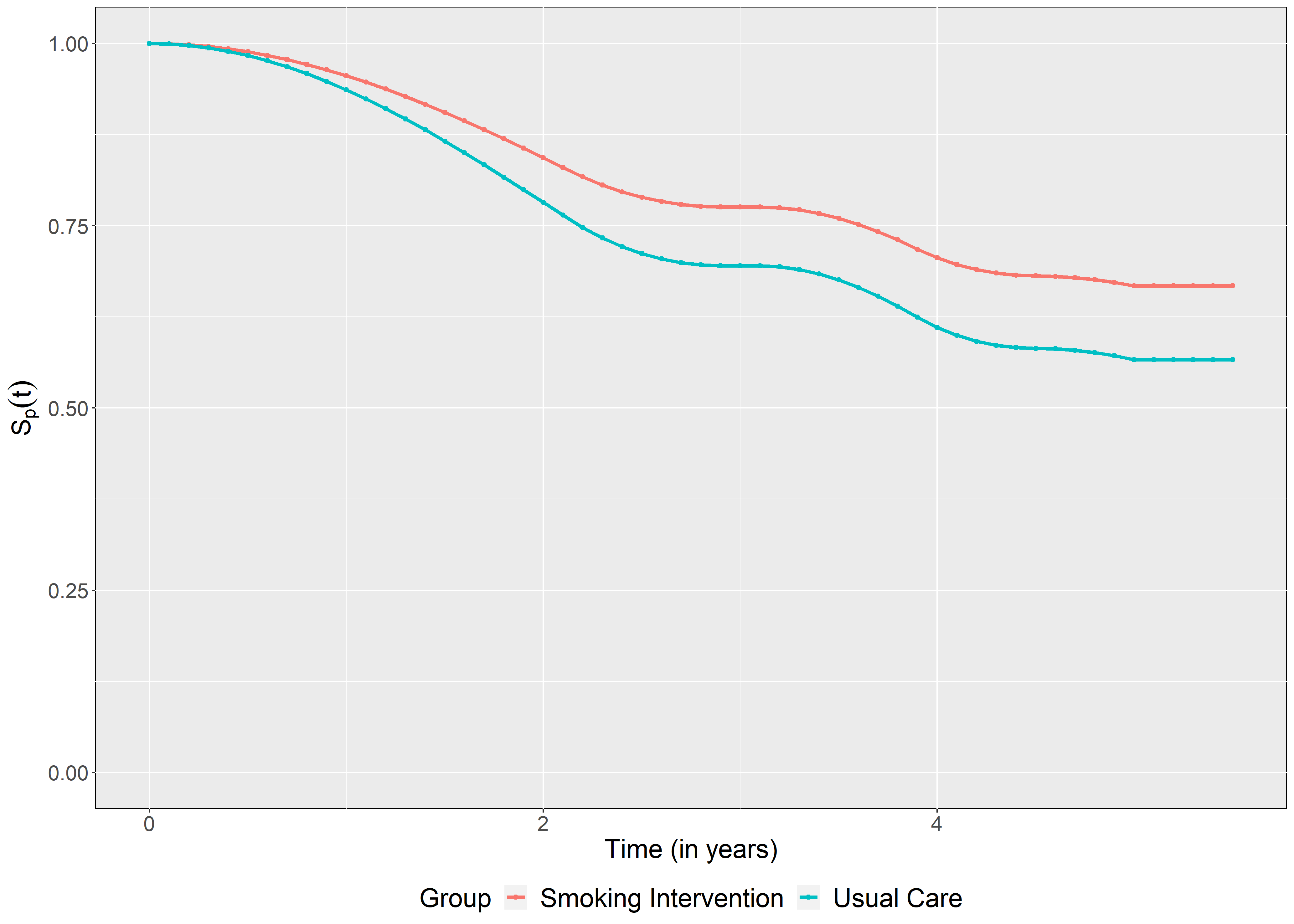}
\caption{Estimated population survival function graph for our proposed model measured at median duration and average number of cigarettes smoked prior to the study enrollment for the smoking cessation data}
\label{Fig1}
\end{figure}

Figure \ref{Fig1} represents estimated survival functions corresponding to our proposed model, and it shows that the group with smoking intervention is likely to have a greater cure rate than the group with usual care.

\section{Concluding remarks} \label{sec:sec6}

In this article, for interval censored data, a flexible family of transformation cure model is proposed that unifies mixture cure model and promotion time cure model. The transformation considered is a Box-Cox transformation taken on the population survival function, and is linked to incidence part and latency part of the cure model. The transformation parameter $\alpha$ provides flexible characteristics to the BCTM, thereby, giving rise to a flexible family of cure models. To introduce non-homogeneity with respect to time-to-events for susceptible subjects, a proportional hazards framework is considered where the baseline hazard function is captured in a distribution-free or non-parametric manner by incorporating a piecewise linear function similar to smoothing splines. Estimation of the model parameters is carried out by implementing the expectation-maximization algorithm to accommodate the missing cure status present in the data. It is to be noted that all parameters including $\alpha$ are considered simultaneously in the  maximization step of the EM algorithm as opposed to applying the profile-likelihood technique considered by authors in previous literature. The performance of our proposed method is studied numerically, and established through a simulation study under various parameter settings (with respect to true $\alpha=0.0, 0.5$ and $1.0$) and sample size ($n=200$ and $400$) consideration, and an application to real-life interval censored data obtained from a smoking cessation study.                

The estimation results obtained from the simulation study are overall satisfactory. The parameter estimates obtained are accurate, consistent with relatively lesser SEs and RMSEs with CPs being closer to the nominal level of 95\% for most of the parameters. The parameter estimate of $\alpha$ shows lesser accuracy, higher SE and RMSE, with CP showing under-coverage for most of the settings. The flatness of the likelihood function surface with respect to $\alpha$ is considered as accountable for this issue. However, simultaneous-maximization approach is found to be $5-10$ times faster than previously used profile-likelihood based approach to estimate $\alpha$ even for basic profile likelihood search range of $\alpha \in \{0.0, 0.1, \dots, 1.0\}$. The SE corresponding to $\psi_B$ is found to be relatively higher when compared with other $\psi_b, b=0, \dots, B-1$. This problem is attributed to lesser precision of estimation of $\psi_B$ owing to greater Euclidean distance between $\tau_{B-1}$ and $\tau_B$, and approximating the true exponential baseline hazard function in $[0, \infty)$ with PLA in the interval $[\tau_0, \tau_B]$. Due to nested nature of the BCTM incorporating PLA in its framework, increasing $B$ is associated with higher likelihood value (even though marginal). Analyzing the real-life example of smoking cessation data with our proposed method, the BCTM with PLA ($B=6$) provides the highest maximized log-likelihood value with the lowest AIC, thereby providing the best fit to the interval censored data. The best fitted BCTM model converges to the MCM model since ML estimate of $\alpha=1$ for the smoking cessation data. An estimated population survival function reflects the presence of significant cure rates in the data, and that the smoking intervention group performs better than the usual care group of subjects.

As a potential future work, it is of great interest to extend the current framework to accommodate covariates with measurement error \cite{barui2020semiparametric}. In this context, it may be also be worth exploring the use of machine learning techniques in the incidence part to capture non-linearity in the data, which may result in improved predictive accuracy of cure \cite{PalSMMR23}. We are currently working on these problems and hope to report the findings in future manuscripts.

\section*{Acknowledgment}

Suvra Pal's research was supported by the National Institute Of General Medical Sciences of the National Institutes of Health under Award Number R15GM150091. The content is solely the responsibility of the author and does not necessarily represent the official views of the National Institutes of Health.

\section*{Conflict of Interest}

The authors have declared no conflict of interest.

\section*{Data Availability Statement}

The R codes for the data generation and the EM algorithm are available in the Supplemental Material.

\bibliographystyle{apacite}
\bibliography{Ref}

\begin{thebibliography}{}

\bibitem [\protect \citeauthoryear {%
Aselisewine%
\ \BBA {} Pal%
}{%
Aselisewine%
\ \BBA {} Pal%
}{%
{\protect \APACyear {2023}}%
}]{%
Ase23}
\APACinsertmetastar {%
Ase23}%
\begin{APACrefauthors}%
Aselisewine, W.%
\BCBT {}\ \BBA {} Pal, S.%
\end{APACrefauthors}%
\unskip\
\newblock
\APACrefYearMonthDay{2023}{}{}.
\newblock
{\BBOQ}\APACrefatitle {On the integration of decision trees with mixture cure
  model} {On the integration of decision trees with mixture cure model}.{\BBCQ}
\newblock
\APACjournalVolNumPages{Statistics in Medicine}{42}{23}{4111-4127}.
\PrintBackRefs{\CurrentBib}

\bibitem [\protect \citeauthoryear {%
Balakrishnan%
\ \BBA {} Barui%
}{%
Balakrishnan%
\ \BBA {} Barui%
}{%
{\protect \APACyear {2023}}%
}]{%
balakrishnan2023destructive}
\APACinsertmetastar {%
balakrishnan2023destructive}%
\begin{APACrefauthors}%
Balakrishnan, N.%
\BCBT {}\ \BBA {} Barui, S.%
\end{APACrefauthors}%
\unskip\
\newblock
\APACrefYearMonthDay{2023}{}{}.
\newblock
{\BBOQ}\APACrefatitle {Destructive cure models with proportional hazards
  lifetimes and associated likelihood inference} {Destructive cure models with
  proportional hazards lifetimes and associated likelihood inference}.{\BBCQ}
\newblock
\APACjournalVolNumPages{Communications in Statistics: Case Studies, Data
  Analysis and Applications}{9}{1}{16--50}.
\PrintBackRefs{\CurrentBib}

\bibitem [\protect \citeauthoryear {%
Balakrishnan%
, Barui%
\BCBL {}\ \BBA {} Milienos%
}{%
Balakrishnan%
\ \protect \BOthers {.}}{%
{\protect \APACyear {2017}}%
}]{%
balakrishnan2017proportional}
\APACinsertmetastar {%
balakrishnan2017proportional}%
\begin{APACrefauthors}%
Balakrishnan, N.%
, Barui, S.%
\BCBL {}\ \BBA {} Milienos, F.%
\end{APACrefauthors}%
\unskip\
\newblock
\APACrefYearMonthDay{2017}{}{}.
\newblock
{\BBOQ}\APACrefatitle {Proportional hazards under {C}onway--{M}axwell-{P}oisson
  cure rate model and associated inference} {Proportional hazards under
  {C}onway--{M}axwell-{P}oisson cure rate model and associated
  inference}.{\BBCQ}
\newblock
\APACjournalVolNumPages{Statistical Methods in Medical
  Research}{26}{5}{2055--2077}.
\PrintBackRefs{\CurrentBib}

\bibitem [\protect \citeauthoryear {%
Balakrishnan%
, Barui%
\BCBL {}\ \BBA {} Milienos%
}{%
Balakrishnan%
\ \protect \BOthers {.}}{%
{\protect \APACyear {2022}}%
}]{%
Bala-Barui22}
\APACinsertmetastar {%
Bala-Barui22}%
\begin{APACrefauthors}%
Balakrishnan, N.%
, Barui, S.%
\BCBL {}\ \BBA {} Milienos, F.%
\end{APACrefauthors}%
\unskip\
\newblock
\APACrefYearMonthDay{2022}{}{}.
\newblock
{\BBOQ}\APACrefatitle {Piecewise linear approximations of baseline under
  proportional hazards based {COM}-{P}oisson cure models} {Piecewise linear
  approximations of baseline under proportional hazards based {COM}-{P}oisson
  cure models}.{\BBCQ}
\newblock
\APACjournalVolNumPages{Communications in Statistics-Simulation and
  Computation}{}{}{1--26}.
\PrintBackRefs{\CurrentBib}

\bibitem [\protect \citeauthoryear {%
Balakrishnan%
, Koutras%
, Milienos%
\BCBL {}\ \BBA {} Pal%
}{%
Balakrishnan%
\ \protect \BOthers {.}}{%
{\protect \APACyear {2016}}%
}]{%
Bal16b}
\APACinsertmetastar {%
Bal16b}%
\begin{APACrefauthors}%
Balakrishnan, N.%
, Koutras, M\BPBI V.%
, Milienos, F\BPBI S.%
\BCBL {}\ \BBA {} Pal, S.%
\end{APACrefauthors}%
\unskip\
\newblock
\APACrefYearMonthDay{2016}{}{}.
\newblock
{\BBOQ}\APACrefatitle {Piecewise linear approximations for cure rate models and
  associated inferential issues} {Piecewise linear approximations for cure rate
  models and associated inferential issues}.{\BBCQ}
\newblock
\APACjournalVolNumPages{Methodology and Computing in Applied
  Probability}{18}{4}{937--966}.
\PrintBackRefs{\CurrentBib}

\bibitem [\protect \citeauthoryear {%
Balakrishnan%
\ \BBA {} Pal%
}{%
Balakrishnan%
\ \BBA {} Pal%
}{%
{\protect \APACyear {2012}}%
}]{%
Bal12}
\APACinsertmetastar {%
Bal12}%
\begin{APACrefauthors}%
Balakrishnan, N.%
\BCBT {}\ \BBA {} Pal, S.%
\end{APACrefauthors}%
\unskip\
\newblock
\APACrefYearMonthDay{2012}{}{}.
\newblock
{\BBOQ}\APACrefatitle {{EM} algorithm-based likelihood estimation for some cure
  rate models} {{EM} algorithm-based likelihood estimation for some cure rate
  models}.{\BBCQ}
\newblock
\APACjournalVolNumPages{Journal of Statistical Theory and
  Practice}{6}{4}{698--724}.
\PrintBackRefs{\CurrentBib}

\bibitem [\protect \citeauthoryear {%
Balakrishnan%
\ \BBA {} Pal%
}{%
Balakrishnan%
\ \BBA {} Pal%
}{%
{\protect \APACyear {2013}}%
}]{%
Bal13}
\APACinsertmetastar {%
Bal13}%
\begin{APACrefauthors}%
Balakrishnan, N.%
\BCBT {}\ \BBA {} Pal, S.%
\end{APACrefauthors}%
\unskip\
\newblock
\APACrefYearMonthDay{2013}{}{}.
\newblock
{\BBOQ}\APACrefatitle {Lognormal lifetimes and likelihood-based inference for
  flexible cure rate models based on {COM}-{P}oisson family} {Lognormal
  lifetimes and likelihood-based inference for flexible cure rate models based
  on {COM}-{P}oisson family}.{\BBCQ}
\newblock
\APACjournalVolNumPages{Computational Statistics \& Data
  Analysis}{67}{}{41--67}.
\PrintBackRefs{\CurrentBib}

\bibitem [\protect \citeauthoryear {%
Balakrishnan%
\ \BBA {} Pal%
}{%
Balakrishnan%
\ \BBA {} Pal%
}{%
{\protect \APACyear {2015}}%
}]{%
Bal15}
\APACinsertmetastar {%
Bal15}%
\begin{APACrefauthors}%
Balakrishnan, N.%
\BCBT {}\ \BBA {} Pal, S.%
\end{APACrefauthors}%
\unskip\
\newblock
\APACrefYearMonthDay{2015}{}{}.
\newblock
{\BBOQ}\APACrefatitle {An {EM} algorithm for the estimation of parameters of a
  flexible cure rate model with generalized gamma lifetime and model
  discrimination using likelihood-and information-based methods} {An {EM}
  algorithm for the estimation of parameters of a flexible cure rate model with
  generalized gamma lifetime and model discrimination using likelihood-and
  information-based methods}.{\BBCQ}
\newblock
\APACjournalVolNumPages{Computational Statistics}{30}{1}{151--189}.
\PrintBackRefs{\CurrentBib}

\bibitem [\protect \citeauthoryear {%
Balakrishnan%
\ \BBA {} Pal%
}{%
Balakrishnan%
\ \BBA {} Pal%
}{%
{\protect \APACyear {2016}}%
}]{%
Bal16}
\APACinsertmetastar {%
Bal16}%
\begin{APACrefauthors}%
Balakrishnan, N.%
\BCBT {}\ \BBA {} Pal, S.%
\end{APACrefauthors}%
\unskip\
\newblock
\APACrefYearMonthDay{2016}{}{}.
\newblock
{\BBOQ}\APACrefatitle {Expectation maximization-based likelihood inference for
  flexible cure rate models with {W}eibull lifetimes} {Expectation
  maximization-based likelihood inference for flexible cure rate models with
  {W}eibull lifetimes}.{\BBCQ}
\newblock
\APACjournalVolNumPages{Statistical Methods in Medical
  Research}{25}{4}{1535--1563}.
\PrintBackRefs{\CurrentBib}

\bibitem [\protect \citeauthoryear {%
Barui%
\ \BBA {} Yi%
}{%
Barui%
\ \BBA {} Yi%
}{%
{\protect \APACyear {2020}}%
}]{%
barui2020semiparametric}
\APACinsertmetastar {%
barui2020semiparametric}%
\begin{APACrefauthors}%
Barui, S.%
\BCBT {}\ \BBA {} Yi, Y\BPBI G.%
\end{APACrefauthors}%
\unskip\
\newblock
\APACrefYearMonthDay{2020}{}{}.
\newblock
{\BBOQ}\APACrefatitle {Semiparametric methods for survival data with
  measurement error under additive hazards cure rate models} {Semiparametric
  methods for survival data with measurement error under additive hazards cure
  rate models}.{\BBCQ}
\newblock
\APACjournalVolNumPages{Lifetime {D}ata {A}nalysis}{26}{3}{421--450}.
\PrintBackRefs{\CurrentBib}

\bibitem [\protect \citeauthoryear {%
Berkson%
\ \BBA {} Gage%
}{%
Berkson%
\ \BBA {} Gage%
}{%
{\protect \APACyear {1952}}%
}]{%
Ber52}
\APACinsertmetastar {%
Ber52}%
\begin{APACrefauthors}%
Berkson, J.%
\BCBT {}\ \BBA {} Gage, R\BPBI P.%
\end{APACrefauthors}%
\unskip\
\newblock
\APACrefYearMonthDay{1952}{}{}.
\newblock
{\BBOQ}\APACrefatitle {Survival curve for cancer patients following treatment}
  {Survival curve for cancer patients following treatment}.{\BBCQ}
\newblock
\APACjournalVolNumPages{Journal of the American Statistical
  Association}{47}{259}{501--515}.
\PrintBackRefs{\CurrentBib}

\bibitem [\protect \citeauthoryear {%
Boag%
}{%
Boag%
}{%
{\protect \APACyear {1949}}%
}]{%
Boa49}
\APACinsertmetastar {%
Boa49}%
\begin{APACrefauthors}%
Boag, J\BPBI W.%
\end{APACrefauthors}%
\unskip\
\newblock
\APACrefYearMonthDay{1949}{}{}.
\newblock
{\BBOQ}\APACrefatitle {Maximum likelihood estimates of the proportion of
  patients cured by cancer therapy} {Maximum likelihood estimates of the
  proportion of patients cured by cancer therapy}.{\BBCQ}
\newblock
\APACjournalVolNumPages{Journal of the Royal Statistical Society. Series B
  (Methodological)}{11}{1}{15--53}.
\PrintBackRefs{\CurrentBib}

\bibitem [\protect \citeauthoryear {%
Chen%
\ \BBA {} Ibrahim%
}{%
Chen%
\ \BBA {} Ibrahim%
}{%
{\protect \APACyear {2001}}%
}]{%
Che01}
\APACinsertmetastar {%
Che01}%
\begin{APACrefauthors}%
Chen, M\BPBI H.%
\BCBT {}\ \BBA {} Ibrahim, J\BPBI G.%
\end{APACrefauthors}%
\unskip\
\newblock
\APACrefYearMonthDay{2001}{}{}.
\newblock
{\BBOQ}\APACrefatitle {Maximum likelihood methods for cure rate models with
  missing covariates} {Maximum likelihood methods for cure rate models with
  missing covariates}.{\BBCQ}
\newblock
\APACjournalVolNumPages{Biometrics}{57}{1}{43--52}.
\PrintBackRefs{\CurrentBib}

\bibitem [\protect \citeauthoryear {%
Farewell%
}{%
Farewell%
}{%
{\protect \APACyear {1982}}%
}]{%
Far82}
\APACinsertmetastar {%
Far82}%
\begin{APACrefauthors}%
Farewell, V\BPBI T.%
\end{APACrefauthors}%
\unskip\
\newblock
\APACrefYearMonthDay{1982}{}{}.
\newblock
{\BBOQ}\APACrefatitle {The use of mixture models for the analysis of survival
  data with long-term survivors} {The use of mixture models for the analysis of
  survival data with long-term survivors}.{\BBCQ}
\newblock
\APACjournalVolNumPages{Biometrics}{}{}{1041--1046}.
\PrintBackRefs{\CurrentBib}

\bibitem [\protect \citeauthoryear {%
Felizzi%
, Paracha%
, P{\"o}hlmann%
\BCBL {}\ \BBA {} Ray%
}{%
Felizzi%
\ \protect \BOthers {.}}{%
{\protect \APACyear {2021}}%
}]{%
felizzi2021mixture}
\APACinsertmetastar {%
felizzi2021mixture}%
\begin{APACrefauthors}%
Felizzi, F.%
, Paracha, N.%
, P{\"o}hlmann, J.%
\BCBL {}\ \BBA {} Ray, J.%
\end{APACrefauthors}%
\unskip\
\newblock
\APACrefYearMonthDay{2021}{}{}.
\newblock
{\BBOQ}\APACrefatitle {Mixture cure models in oncology: a tutorial and
  practical guidance} {Mixture cure models in oncology: a tutorial and
  practical guidance}.{\BBCQ}
\newblock
\APACjournalVolNumPages{PharmacoEconomics-Open}{5}{}{143--155}.
\PrintBackRefs{\CurrentBib}

\bibitem [\protect \citeauthoryear {%
Hoang%
, Pal%
, Liu%
, Senkowsky%
\BCBL {}\ \BBA {} Tang%
}{%
Hoang%
\ \protect \BOthers {.}}{%
{\protect \APACyear {2023}}%
}]{%
Hoang23}
\APACinsertmetastar {%
Hoang23}%
\begin{APACrefauthors}%
Hoang, L.%
, Pal, S.%
, Liu, Z.%
, Senkowsky, J.%
\BCBL {}\ \BBA {} Tang, L.%
\end{APACrefauthors}%
\unskip\
\newblock
\APACrefYearMonthDay{2023}{}{}.
\newblock
{\BBOQ}\APACrefatitle {A time‐dependent survival analysis for early prognosis
  of chronic wounds by monitoring wound alkalinity} {A time‐dependent
  survival analysis for early prognosis of chronic wounds by monitoring wound
  alkalinity}.{\BBCQ}
\newblock
\APACjournalVolNumPages{International Wound Journal}{20}{5}{1459--1475}.
\PrintBackRefs{\CurrentBib}

\bibitem [\protect \citeauthoryear {%
Ibrahim%
, Chen%
\BCBL {}\ \BBA {} Sinha%
}{%
Ibrahim%
\ \protect \BOthers {.}}{%
{\protect \APACyear {2001}}%
}]{%
Ibr01}
\APACinsertmetastar {%
Ibr01}%
\begin{APACrefauthors}%
Ibrahim, J\BPBI G.%
, Chen, M\BPBI H.%
\BCBL {}\ \BBA {} Sinha, D.%
\end{APACrefauthors}%
\unskip\
\newblock
\APACrefYear{2001}.
\newblock
\APACrefbtitle {Bayesian {S}urvival {A}nalysis} {Bayesian {S}urvival
  {A}nalysis}.
\newblock
\APACaddressPublisher{New York}{Springer}.
\PrintBackRefs{\CurrentBib}

\bibitem [\protect \citeauthoryear {%
Kim%
\ \BBA {} Jhun%
}{%
Kim%
\ \BBA {} Jhun%
}{%
{\protect \APACyear {2008}}%
}]{%
Kim08}
\APACinsertmetastar {%
Kim08}%
\begin{APACrefauthors}%
Kim, Y\BPBI J.%
\BCBT {}\ \BBA {} Jhun, M.%
\end{APACrefauthors}%
\unskip\
\newblock
\APACrefYearMonthDay{2008}{}{}.
\newblock
{\BBOQ}\APACrefatitle {Cure rate model with interval censored data} {Cure rate
  model with interval censored data}.{\BBCQ}
\newblock
\APACjournalVolNumPages{Statistics in Medicine}{27}{1}{3--14}.
\PrintBackRefs{\CurrentBib}

\bibitem [\protect \citeauthoryear {%
Koutras%
\ \BBA {} Milienos%
}{%
Koutras%
\ \BBA {} Milienos%
}{%
{\protect \APACyear {2017}}%
}]{%
Kou17}
\APACinsertmetastar {%
Kou17}%
\begin{APACrefauthors}%
Koutras, M\BPBI V.%
\BCBT {}\ \BBA {} Milienos, F\BPBI S.%
\end{APACrefauthors}%
\unskip\
\newblock
\APACrefYearMonthDay{2017}{}{}.
\newblock
{\BBOQ}\APACrefatitle {A flexible family of transformation cure rate models} {A
  flexible family of transformation cure rate models}.{\BBCQ}
\newblock
\APACjournalVolNumPages{Statistics in Medicine}{36}{16}{2559-2575}.
\PrintBackRefs{\CurrentBib}

\bibitem [\protect \citeauthoryear {%
Kuk%
\ \BBA {} Chen%
}{%
Kuk%
\ \BBA {} Chen%
}{%
{\protect \APACyear {1992}}%
}]{%
Kuk92}
\APACinsertmetastar {%
Kuk92}%
\begin{APACrefauthors}%
Kuk, A\BPBI Y\BPBI C.%
\BCBT {}\ \BBA {} Chen, C\BPBI H.%
\end{APACrefauthors}%
\unskip\
\newblock
\APACrefYearMonthDay{1992}{}{}.
\newblock
{\BBOQ}\APACrefatitle {A mixture model combining logistic regression with
  proportional hazards regression} {A mixture model combining logistic
  regression with proportional hazards regression}.{\BBCQ}
\newblock
\APACjournalVolNumPages{Biometrika}{79}{3}{531--541}.
\PrintBackRefs{\CurrentBib}

\bibitem [\protect \citeauthoryear {%
Larson%
\ \BBA {} Dinse%
}{%
Larson%
\ \BBA {} Dinse%
}{%
{\protect \APACyear {1985}}%
}]{%
Lar85}
\APACinsertmetastar {%
Lar85}%
\begin{APACrefauthors}%
Larson, M\BPBI G.%
\BCBT {}\ \BBA {} Dinse, G\BPBI E.%
\end{APACrefauthors}%
\unskip\
\newblock
\APACrefYearMonthDay{1985}{}{}.
\newblock
{\BBOQ}\APACrefatitle {A mixture model for the regression analysis of competing
  risks data} {A mixture model for the regression analysis of competing risks
  data}.{\BBCQ}
\newblock
\APACjournalVolNumPages{Applied Statistics}{}{}{201--211}.
\PrintBackRefs{\CurrentBib}

\bibitem [\protect \citeauthoryear {%
Lindsey%
\ \BBA {} Ryan%
}{%
Lindsey%
\ \BBA {} Ryan%
}{%
{\protect \APACyear {1998}}%
}]{%
lindsey1998methods}
\APACinsertmetastar {%
lindsey1998methods}%
\begin{APACrefauthors}%
Lindsey, J\BPBI C.%
\BCBT {}\ \BBA {} Ryan, L\BPBI M.%
\end{APACrefauthors}%
\unskip\
\newblock
\APACrefYearMonthDay{1998}{}{}.
\newblock
{\BBOQ}\APACrefatitle {Methods for interval-censored data} {Methods for
  interval-censored data}.{\BBCQ}
\newblock
\APACjournalVolNumPages{Statistics in Medicine}{17}{2}{219--238}.
\PrintBackRefs{\CurrentBib}

\bibitem [\protect \citeauthoryear {%
Lo%
, Taylor%
, McBride%
\BCBL {}\ \BBA {} Withers%
}{%
Lo%
\ \protect \BOthers {.}}{%
{\protect \APACyear {1993}}%
}]{%
Lo93}
\APACinsertmetastar {%
Lo93}%
\begin{APACrefauthors}%
Lo, Y\BHBI C.%
, Taylor, J\BPBI M.%
, McBride, W\BPBI H.%
\BCBL {}\ \BBA {} Withers, H\BPBI R.%
\end{APACrefauthors}%
\unskip\
\newblock
\APACrefYearMonthDay{1993}{}{}.
\newblock
{\BBOQ}\APACrefatitle {The effect of fractionated doses of radiation on mouse
  spinal cord} {The effect of fractionated doses of radiation on mouse spinal
  cord}.{\BBCQ}
\newblock
\APACjournalVolNumPages{International Journal of Radiation Oncology, Biology,
  Physics}{27}{2}{309--317}.
\PrintBackRefs{\CurrentBib}

\bibitem [\protect \citeauthoryear {%
Ma%
}{%
Ma%
}{%
{\protect \APACyear {2010}}%
}]{%
Ma10}
\APACinsertmetastar {%
Ma10}%
\begin{APACrefauthors}%
Ma, S.%
\end{APACrefauthors}%
\unskip\
\newblock
\APACrefYearMonthDay{2010}{}{}.
\newblock
{\BBOQ}\APACrefatitle {Mixed case interval censored data with a cured subgroup}
  {Mixed case interval censored data with a cured subgroup}.{\BBCQ}
\newblock
\APACjournalVolNumPages{Statistica Sinica}{20}{3}{1165--1181}.
\PrintBackRefs{\CurrentBib}

\bibitem [\protect \citeauthoryear {%
Majakwara%
\ \BBA {} Pal%
}{%
Majakwara%
\ \BBA {} Pal%
}{%
{\protect \APACyear {2019}}%
}]{%
Majak19}
\APACinsertmetastar {%
Majak19}%
\begin{APACrefauthors}%
Majakwara, J.%
\BCBT {}\ \BBA {} Pal, S.%
\end{APACrefauthors}%
\unskip\
\newblock
\APACrefYearMonthDay{2019}{}{}.
\newblock
{\BBOQ}\APACrefatitle {On some inferential issues for the destructive
  {COM}-{P}oisson-generalized gamma regression cure rate model} {On some
  inferential issues for the destructive {COM}-{P}oisson-generalized gamma
  regression cure rate model}.{\BBCQ}
\newblock
\APACjournalVolNumPages{Communications in Statistics - Simulation and
  Computation}{48}{10}{3118--3142}.
\PrintBackRefs{\CurrentBib}

\bibitem [\protect \citeauthoryear {%
Milienos%
}{%
Milienos%
}{%
{\protect \APACyear {2022}}%
}]{%
Mil22}
\APACinsertmetastar {%
Mil22}%
\begin{APACrefauthors}%
Milienos, F\BPBI S.%
\end{APACrefauthors}%
\unskip\
\newblock
\APACrefYearMonthDay{2022}{}{}.
\newblock
{\BBOQ}\APACrefatitle {On a reparameterization of a flexible family of cure
  models} {On a reparameterization of a flexible family of cure models}.{\BBCQ}
\newblock
\APACjournalVolNumPages{Statistics in Medicine}{41}{}{4091--4111}.
\PrintBackRefs{\CurrentBib}

\bibitem [\protect \citeauthoryear {%
Murray%
\ \protect \BOthers {.}}{%
Murray%
\ \protect \BOthers {.}}{%
{\protect \APACyear {1998}}%
}]{%
murray1998effects}
\APACinsertmetastar {%
murray1998effects}%
\begin{APACrefauthors}%
Murray, R\BPBI P.%
, Anthonisen, N\BPBI R.%
, Connett, J\BPBI E.%
, Wise, R\BPBI A.%
, Lindgren, P\BPBI G.%
, Greene, P\BPBI G.%
\BDBL {}others%
\end{APACrefauthors}%
\unskip\
\newblock
\APACrefYearMonthDay{1998}{}{}.
\newblock
{\BBOQ}\APACrefatitle {Effects of multiple attempts to quit smoking and
  relapses to smoking on pulmonary function} {Effects of multiple attempts to
  quit smoking and relapses to smoking on pulmonary function}.{\BBCQ}
\newblock
\APACjournalVolNumPages{Journal of Clinical Epidemiology}{51}{12}{1317--1326}.
\PrintBackRefs{\CurrentBib}

\bibitem [\protect \citeauthoryear {%
Othus%
, Barlogie%
, LeBlanc%
\BCBL {}\ \BBA {} Crowley%
}{%
Othus%
\ \protect \BOthers {.}}{%
{\protect \APACyear {2012}}%
}]{%
othus2012cure}
\APACinsertmetastar {%
othus2012cure}%
\begin{APACrefauthors}%
Othus, M.%
, Barlogie, B.%
, LeBlanc, M\BPBI L.%
\BCBL {}\ \BBA {} Crowley, J\BPBI J.%
\end{APACrefauthors}%
\unskip\
\newblock
\APACrefYearMonthDay{2012}{}{}.
\newblock
{\BBOQ}\APACrefatitle {Cure models as a useful statistical tool for analyzing
  survival} {Cure models as a useful statistical tool for analyzing
  survival}.{\BBCQ}
\newblock
\APACjournalVolNumPages{Clinical Cancer Research}{18}{14}{3731--3736}.
\PrintBackRefs{\CurrentBib}

\bibitem [\protect \citeauthoryear {%
Overman%
\ \BBA {} Pal%
}{%
Overman%
\ \BBA {} Pal%
}{%
{\protect \APACyear {2022}}%
}]{%
Tom21}
\APACinsertmetastar {%
Tom21}%
\begin{APACrefauthors}%
Overman, T.%
\BCBT {}\ \BBA {} Pal, S.%
\end{APACrefauthors}%
\unskip\
\newblock
\APACrefYearMonthDay{2022}{}{}.
\newblock
{\BBOQ}\APACrefatitle {Statistical Tools and Techniques in Modeling Survival
  Data} {Statistical tools and techniques in modeling survival data}.{\BBCQ}
\newblock
\BIn{} E\BPBI E.~Goldwyn, S.~Ganzell\BCBL {}\ \BBA {} A.~Wootton\ (\BEDS),
  \APACrefbtitle {Mathematics Research for the Beginning Student, Volume 2:
  Accessible Projects for Students After Calculus} {Mathematics research for
  the beginning student, volume 2: Accessible projects for students after
  calculus}\ (\BPGS\ 75--99).
\newblock
\APACaddressPublisher{Cham}{Springer International Publishing}.
\PrintBackRefs{\CurrentBib}

\bibitem [\protect \citeauthoryear {%
Pal%
\ \BBA {} Aselisewine%
}{%
Pal%
\ \BBA {} Aselisewine%
}{%
{\protect \APACyear {2023}}%
}]{%
PalSVM23}
\APACinsertmetastar {%
PalSVM23}%
\begin{APACrefauthors}%
Pal, S.%
\BCBT {}\ \BBA {} Aselisewine, W.%
\end{APACrefauthors}%
\unskip\
\newblock
\APACrefYearMonthDay{2023}{}{}.
\newblock
{\BBOQ}\APACrefatitle {A semiparametric promotion time cure model with support
  vector machine} {A semiparametric promotion time cure model with support
  vector machine}.{\BBCQ}
\newblock
\APACjournalVolNumPages{Annals of Applied Statistics}{17}{3}{2680-2699}.
\PrintBackRefs{\CurrentBib}

\bibitem [\protect \citeauthoryear {%
Pal%
\ \BBA {} Balakrishnan%
}{%
Pal%
\ \BBA {} Balakrishnan%
}{%
{\protect \APACyear {2016}}%
}]{%
Pal16}
\APACinsertmetastar {%
Pal16}%
\begin{APACrefauthors}%
Pal, S.%
\BCBT {}\ \BBA {} Balakrishnan, N.%
\end{APACrefauthors}%
\unskip\
\newblock
\APACrefYearMonthDay{2016}{}{}.
\newblock
{\BBOQ}\APACrefatitle {Destructive negative binomial cure rate model and
  {EM}-based likelihood inference under {W}eibull lifetime} {Destructive
  negative binomial cure rate model and {EM}-based likelihood inference under
  {W}eibull lifetime}.{\BBCQ}
\newblock
\APACjournalVolNumPages{Statistics \& Probability Letters}{116}{}{9--20}.
\PrintBackRefs{\CurrentBib}

\bibitem [\protect \citeauthoryear {%
Pal%
\ \BBA {} Balakrishnan%
}{%
Pal%
\ \BBA {} Balakrishnan%
}{%
{\protect \APACyear {2017}}%
{\protect \APACexlab {{\protect \BCnt {1}}}}}]{%
Pal17b}
\APACinsertmetastar {%
Pal17b}%
\begin{APACrefauthors}%
Pal, S.%
\BCBT {}\ \BBA {} Balakrishnan, N.%
\end{APACrefauthors}%
\unskip\
\newblock
\APACrefYearMonthDay{2017{\protect \BCnt {1}}}{}{}.
\newblock
{\BBOQ}\APACrefatitle {An {EM} type estimation procedure for the destructive
  exponentially weighted {P}oisson regression cure model under generalized
  gamma lifetime} {An {EM} type estimation procedure for the destructive
  exponentially weighted {P}oisson regression cure model under generalized
  gamma lifetime}.{\BBCQ}
\newblock
\APACjournalVolNumPages{Journal of Statistical Computation and
  Simulation}{87}{6}{1107--1129}.
\PrintBackRefs{\CurrentBib}

\bibitem [\protect \citeauthoryear {%
Pal%
\ \BBA {} Balakrishnan%
}{%
Pal%
\ \BBA {} Balakrishnan%
}{%
{\protect \APACyear {2017}}%
{\protect \APACexlab {{\protect \BCnt {2}}}}}]{%
Pal17c}
\APACinsertmetastar {%
Pal17c}%
\begin{APACrefauthors}%
Pal, S.%
\BCBT {}\ \BBA {} Balakrishnan, N.%
\end{APACrefauthors}%
\unskip\
\newblock
\APACrefYearMonthDay{2017{\protect \BCnt {2}}}{}{}.
\newblock
{\BBOQ}\APACrefatitle {Likelihood inference for {COM}-{P}oisson cure rate model
  with interval-censored data and {W}eibull lifetimes} {Likelihood inference
  for {COM}-{P}oisson cure rate model with interval-censored data and {W}eibull
  lifetimes}.{\BBCQ}
\newblock
\APACjournalVolNumPages{Statistical Methods in Medical
  Research}{26}{5}{2093-2113}.
\PrintBackRefs{\CurrentBib}

\bibitem [\protect \citeauthoryear {%
Pal%
\ \BBA {} Balakrishnan%
}{%
Pal%
\ \BBA {} Balakrishnan%
}{%
{\protect \APACyear {2017}}%
{\protect \APACexlab {{\protect \BCnt {3}}}}}]{%
Pal17a}
\APACinsertmetastar {%
Pal17a}%
\begin{APACrefauthors}%
Pal, S.%
\BCBT {}\ \BBA {} Balakrishnan, N.%
\end{APACrefauthors}%
\unskip\
\newblock
\APACrefYearMonthDay{2017{\protect \BCnt {3}}}{}{}.
\newblock
{\BBOQ}\APACrefatitle {Likelihood inference for the destructive exponentially
  weighted {P}oisson cure rate model with {W}eibull lifetime and an application
  to melanoma data} {Likelihood inference for the destructive exponentially
  weighted {P}oisson cure rate model with {W}eibull lifetime and an application
  to melanoma data}.{\BBCQ}
\newblock
\APACjournalVolNumPages{Computational Statistics}{32}{2}{429--449}.
\PrintBackRefs{\CurrentBib}

\bibitem [\protect \citeauthoryear {%
Pal%
\ \BBA {} Balakrishnan%
}{%
Pal%
\ \BBA {} Balakrishnan%
}{%
{\protect \APACyear {2018}}%
}]{%
Pal18c}
\APACinsertmetastar {%
Pal18c}%
\begin{APACrefauthors}%
Pal, S.%
\BCBT {}\ \BBA {} Balakrishnan, N.%
\end{APACrefauthors}%
\unskip\
\newblock
\APACrefYearMonthDay{2018}{}{}.
\newblock
{\BBOQ}\APACrefatitle {Expectation maximization algorithm for {B}ox--{C}ox
  transformation cure rate model and assessment of model misspecification under
  {W}eibull lifetimes} {Expectation maximization algorithm for {B}ox--{C}ox
  transformation cure rate model and assessment of model misspecification under
  {W}eibull lifetimes}.{\BBCQ}
\newblock
\APACjournalVolNumPages{IEEE Journal of Biomedical and Health
  Informatics}{22}{3}{926-934}.
\PrintBackRefs{\CurrentBib}

\bibitem [\protect \citeauthoryear {%
Pal%
, Barui%
, Davies%
\BCBL {}\ \BBA {} Mishra%
}{%
Pal%
\ \protect \BOthers {.}}{%
{\protect \APACyear {2022}}%
}]{%
pal2022stochastic}
\APACinsertmetastar {%
pal2022stochastic}%
\begin{APACrefauthors}%
Pal, S.%
, Barui, S.%
, Davies, K.%
\BCBL {}\ \BBA {} Mishra, N.%
\end{APACrefauthors}%
\unskip\
\newblock
\APACrefYearMonthDay{2022}{}{}.
\newblock
{\BBOQ}\APACrefatitle {A Stochastic Version of the EM Algorithm for Mixture
  Cure Model with Exponentiated Weibull Family of Lifetimes} {A stochastic
  version of the em algorithm for mixture cure model with exponentiated weibull
  family of lifetimes}.{\BBCQ}
\newblock
\APACjournalVolNumPages{Journal of Statistical Theory and Practice}{16}{3}{48}.
\PrintBackRefs{\CurrentBib}

\bibitem [\protect \citeauthoryear {%
Pal%
, Majakwara%
\BCBL {}\ \BBA {} Balakrishnan%
}{%
Pal%
\ \protect \BOthers {.}}{%
{\protect \APACyear {2018}}%
}]{%
Pal18}
\APACinsertmetastar {%
Pal18}%
\begin{APACrefauthors}%
Pal, S.%
, Majakwara, J.%
\BCBL {}\ \BBA {} Balakrishnan, N.%
\end{APACrefauthors}%
\unskip\
\newblock
\APACrefYearMonthDay{2018}{}{}.
\newblock
{\BBOQ}\APACrefatitle {An {EM} algorithm for the destructive {COM-P}oisson
  regression cure rate model} {An {EM} algorithm for the destructive
  {COM-P}oisson regression cure rate model}.{\BBCQ}
\newblock
\APACjournalVolNumPages{Metrika}{81}{2}{143--171}.
\PrintBackRefs{\CurrentBib}

\bibitem [\protect \citeauthoryear {%
Pal%
, Peng%
\BCBL {}\ \BBA {} Aselisewine%
}{%
Pal%
, Peng%
\BCBL {}\ \BBA {} Aselisewine%
}{%
{\protect \APACyear {2023}}%
}]{%
Paletal23}
\APACinsertmetastar {%
Paletal23}%
\begin{APACrefauthors}%
Pal, S.%
, Peng, Y.%
\BCBL {}\ \BBA {} Aselisewine, W.%
\end{APACrefauthors}%
\unskip\
\newblock
\APACrefYearMonthDay{2023}{}{}.
\newblock
{\BBOQ}\APACrefatitle {A new approach to modeling the cure rate in the presence
  of interval censored data} {A new approach to modeling the cure rate in the
  presence of interval censored data}.{\BBCQ}
\newblock
\APACjournalVolNumPages{Computational
  Statistics}{DOI:10.1007/s00180-023-01389-7}{}{}.
\PrintBackRefs{\CurrentBib}

\bibitem [\protect \citeauthoryear {%
Pal%
, Peng%
, Aselisewine%
\BCBL {}\ \BBA {} Barui%
}{%
Pal%
, Peng%
, Aselisewine%
\BCBL {}\ \BBA {} Barui%
}{%
{\protect \APACyear {2023}}%
}]{%
PalSMMR23}
\APACinsertmetastar {%
PalSMMR23}%
\begin{APACrefauthors}%
Pal, S.%
, Peng, Y.%
, Aselisewine, W.%
\BCBL {}\ \BBA {} Barui, S.%
\end{APACrefauthors}%
\unskip\
\newblock
\APACrefYearMonthDay{2023}{}{}.
\newblock
{\BBOQ}\APACrefatitle {A support vector machine based cure rate model for
  interval censored data} {A support vector machine based cure rate model for
  interval censored data}.{\BBCQ}
\newblock
\APACjournalVolNumPages{Statistical Methods in Medical Research}{DOI:
  10.1177/09622802231210917}{}{}.
\PrintBackRefs{\CurrentBib}

\bibitem [\protect \citeauthoryear {%
Pal%
\ \BBA {} Roy%
}{%
Pal%
\ \BBA {} Roy%
}{%
{\protect \APACyear {2021}}%
}]{%
PalRoy21}
\APACinsertmetastar {%
PalRoy21}%
\begin{APACrefauthors}%
Pal, S.%
\BCBT {}\ \BBA {} Roy, S.%
\end{APACrefauthors}%
\unskip\
\newblock
\APACrefYearMonthDay{2021}{}{}.
\newblock
{\BBOQ}\APACrefatitle {On the estimation of destructive cure rate model: a new
  study with exponentially weighted {P}oisson competing risks} {On the
  estimation of destructive cure rate model: a new study with exponentially
  weighted {P}oisson competing risks}.{\BBCQ}
\newblock
\APACjournalVolNumPages{Statistica Neerlandica}{75}{3}{324--342}.
\PrintBackRefs{\CurrentBib}

\bibitem [\protect \citeauthoryear {%
Pal%
\ \BBA {} Roy%
}{%
Pal%
\ \BBA {} Roy%
}{%
{\protect \APACyear {2022}}%
}]{%
PalRoy22}
\APACinsertmetastar {%
PalRoy22}%
\begin{APACrefauthors}%
Pal, S.%
\BCBT {}\ \BBA {} Roy, S.%
\end{APACrefauthors}%
\unskip\
\newblock
\APACrefYearMonthDay{2022}{}{}.
\newblock
{\BBOQ}\APACrefatitle {A new non-linear conjugate gradient algorithm for
  destructive cure rate model and a simulation study: illustration with
  negative binomial competing risks} {A new non-linear conjugate gradient
  algorithm for destructive cure rate model and a simulation study:
  illustration with negative binomial competing risks}.{\BBCQ}
\newblock
\APACjournalVolNumPages{Communications in Statistics - Simulation and
  Computation}{51}{11}{6866--6880}.
\PrintBackRefs{\CurrentBib}

\bibitem [\protect \citeauthoryear {%
Pal%
\ \BBA {} Roy%
}{%
Pal%
\ \BBA {} Roy%
}{%
{\protect \APACyear {2023}}%
}]{%
PalRoy23}
\APACinsertmetastar {%
PalRoy23}%
\begin{APACrefauthors}%
Pal, S.%
\BCBT {}\ \BBA {} Roy, S.%
\end{APACrefauthors}%
\unskip\
\newblock
\APACrefYearMonthDay{2023}{}{}.
\newblock
{\BBOQ}\APACrefatitle {On the parameter estimation of {B}ox-{C}ox
  transformation cure model} {On the parameter estimation of {B}ox-{C}ox
  transformation cure model}.{\BBCQ}
\newblock
\APACjournalVolNumPages{Statistics in Medicine}{42}{15}{2600--2618}.
\PrintBackRefs{\CurrentBib}

\bibitem [\protect \citeauthoryear {%
Peng%
\ \BBA {} Taylor%
}{%
Peng%
\ \BBA {} Taylor%
}{%
{\protect \APACyear {2014}}%
}]{%
peng2014cure}
\APACinsertmetastar {%
peng2014cure}%
\begin{APACrefauthors}%
Peng, Y.%
\BCBT {}\ \BBA {} Taylor, J\BPBI M.%
\end{APACrefauthors}%
\unskip\
\newblock
\APACrefYearMonthDay{2014}{}{}.
\newblock
{\BBOQ}\APACrefatitle {Cure models} {Cure models}.{\BBCQ}
\newblock
\APACjournalVolNumPages{Handbook of Survival Analysis}{34}{}{113--134}.
\PrintBackRefs{\CurrentBib}

\bibitem [\protect \citeauthoryear {%
Rodrigues%
, de Castro%
, Cancho%
\BCBL {}\ \BBA {} Balakrishnan%
}{%
Rodrigues%
\ \protect \BOthers {.}}{%
{\protect \APACyear {2009}}%
}]{%
Rod09}
\APACinsertmetastar {%
Rod09}%
\begin{APACrefauthors}%
Rodrigues, J.%
, de Castro, M.%
, Cancho, V\BPBI G.%
\BCBL {}\ \BBA {} Balakrishnan, N.%
\end{APACrefauthors}%
\unskip\
\newblock
\APACrefYearMonthDay{2009}{}{}.
\newblock
{\BBOQ}\APACrefatitle {{COM--P}oisson cure rate survival models and an
  application to a cutaneous melanoma data} {{COM--P}oisson cure rate survival
  models and an application to a cutaneous melanoma data}.{\BBCQ}
\newblock
\APACjournalVolNumPages{Journal of Statistical Planning and
  Inference}{139}{10}{3605--3611}.
\PrintBackRefs{\CurrentBib}

\bibitem [\protect \citeauthoryear {%
Sy%
\ \BBA {} Taylor%
}{%
Sy%
\ \BBA {} Taylor%
}{%
{\protect \APACyear {2000}}%
}]{%
Sy00}
\APACinsertmetastar {%
Sy00}%
\begin{APACrefauthors}%
Sy, J\BPBI P.%
\BCBT {}\ \BBA {} Taylor, J\BPBI M\BPBI G.%
\end{APACrefauthors}%
\unskip\
\newblock
\APACrefYearMonthDay{2000}{}{}.
\newblock
{\BBOQ}\APACrefatitle {Estimation in a {C}ox proportional hazards cure model}
  {Estimation in a {C}ox proportional hazards cure model}.{\BBCQ}
\newblock
\APACjournalVolNumPages{Biometrics}{56}{1}{227--236}.
\PrintBackRefs{\CurrentBib}

\bibitem [\protect \citeauthoryear {%
Sy%
\ \BBA {} Yaylor%
}{%
Sy%
\ \BBA {} Yaylor%
}{%
{\protect \APACyear {2001}}%
}]{%
Sy01}
\APACinsertmetastar {%
Sy01}%
\begin{APACrefauthors}%
Sy, J\BPBI P.%
\BCBT {}\ \BBA {} Yaylor, J\BPBI M\BPBI G.%
\end{APACrefauthors}%
\unskip\
\newblock
\APACrefYearMonthDay{2001}{}{}.
\newblock
{\BBOQ}\APACrefatitle {Standard errors for the {C}ox proportional hazards cure
  model} {Standard errors for the {C}ox proportional hazards cure
  model}.{\BBCQ}
\newblock
\APACjournalVolNumPages{Mathematical and Computer
  Modelling}{33}{12}{1237--1251}.
\PrintBackRefs{\CurrentBib}

\bibitem [\protect \citeauthoryear {%
Treszoks%
\ \BBA {} Pal%
}{%
Treszoks%
\ \BBA {} Pal%
}{%
{\protect \APACyear {2022}}%
}]{%
Jodi22}
\APACinsertmetastar {%
Jodi22}%
\begin{APACrefauthors}%
Treszoks, J.%
\BCBT {}\ \BBA {} Pal, S.%
\end{APACrefauthors}%
\unskip\
\newblock
\APACrefYearMonthDay{2022}{}{}.
\newblock
{\BBOQ}\APACrefatitle {A destructive shifted {P}oisson cure model for interval
  censored data and an efficient estimation algorithm} {A destructive shifted
  {P}oisson cure model for interval censored data and an efficient estimation
  algorithm}.{\BBCQ}
\newblock
\APACjournalVolNumPages{Communications in Statistics-Simulation and
  Computation}{DOI:10.1080/03610918.2022.2067876}{}{}.
\PrintBackRefs{\CurrentBib}

\bibitem [\protect \citeauthoryear {%
Treszoks%
\ \BBA {} Pal%
}{%
Treszoks%
\ \BBA {} Pal%
}{%
{\protect \APACyear {2023}}%
}]{%
JodiSIM23}
\APACinsertmetastar {%
JodiSIM23}%
\begin{APACrefauthors}%
Treszoks, J.%
\BCBT {}\ \BBA {} Pal, S.%
\end{APACrefauthors}%
\unskip\
\newblock
\APACrefYearMonthDay{2023}{}{}.
\newblock
{\BBOQ}\APACrefatitle {On the estimation of interval censored destructive
  negative binomial cure model} {On the estimation of interval censored
  destructive negative binomial cure model}.{\BBCQ}
\newblock
\APACjournalVolNumPages{Statistics in Medicine}{DOI:10.1002/sim.9904}{}{}.
\PrintBackRefs{\CurrentBib}

\bibitem [\protect \citeauthoryear {%
Tsodikov%
, Ibrahim%
\BCBL {}\ \BBA {} Yakovlev%
}{%
Tsodikov%
\ \protect \BOthers {.}}{%
{\protect \APACyear {2003}}%
}]{%
Tso03}
\APACinsertmetastar {%
Tso03}%
\begin{APACrefauthors}%
Tsodikov, A\BPBI D.%
, Ibrahim, J\BPBI G.%
\BCBL {}\ \BBA {} Yakovlev, A\BPBI Y.%
\end{APACrefauthors}%
\unskip\
\newblock
\APACrefYearMonthDay{2003}{}{}.
\newblock
{\BBOQ}\APACrefatitle {Estimating cure rates from survival data: an alternative
  to two-component mixture models} {Estimating cure rates from survival data:
  an alternative to two-component mixture models}.{\BBCQ}
\newblock
\APACjournalVolNumPages{Journal of the American Statistical
  Association}{98}{464}{1063--1078}.
\PrintBackRefs{\CurrentBib}

\bibitem [\protect \citeauthoryear {%
Wang%
\ \BBA {} Pal%
}{%
Wang%
\ \BBA {} Pal%
}{%
{\protect \APACyear {2022}}%
}]{%
Wang22}
\APACinsertmetastar {%
Wang22}%
\begin{APACrefauthors}%
Wang, P.%
\BCBT {}\ \BBA {} Pal, S.%
\end{APACrefauthors}%
\unskip\
\newblock
\APACrefYearMonthDay{2022}{}{}.
\newblock
{\BBOQ}\APACrefatitle {A two‐way flexible generalized gamma transformation
  cure rate model} {A two‐way flexible generalized gamma transformation cure
  rate model}.{\BBCQ}
\newblock
\APACjournalVolNumPages{Statistics in Medicine}{41}{13}{2427--2447}.
\PrintBackRefs{\CurrentBib}

\bibitem [\protect \citeauthoryear {%
Wiangnak%
\ \BBA {} Pal%
}{%
Wiangnak%
\ \BBA {} Pal%
}{%
{\protect \APACyear {2018}}%
}]{%
Wia18}
\APACinsertmetastar {%
Wia18}%
\begin{APACrefauthors}%
Wiangnak, P.%
\BCBT {}\ \BBA {} Pal, S.%
\end{APACrefauthors}%
\unskip\
\newblock
\APACrefYearMonthDay{2018}{}{}.
\newblock
{\BBOQ}\APACrefatitle {Gamma lifetimes and associated inference for
  interval-censored cure rate model with {COM}--{P}oisson competing cause}
  {Gamma lifetimes and associated inference for interval-censored cure rate
  model with {COM}--{P}oisson competing cause}.{\BBCQ}
\newblock
\APACjournalVolNumPages{Communications in Statistics - Theory and
  Methods}{47}{6}{1491-1509}.
\PrintBackRefs{\CurrentBib}

\bibitem [\protect \citeauthoryear {%
Xiang%
, Ma%
\BCBL {}\ \BBA {} Yau%
}{%
Xiang%
\ \protect \BOthers {.}}{%
{\protect \APACyear {2011}}%
}]{%
Xia11}
\APACinsertmetastar {%
Xia11}%
\begin{APACrefauthors}%
Xiang, L.%
, Ma, X.%
\BCBL {}\ \BBA {} Yau, K\BPBI K\BPBI W.%
\end{APACrefauthors}%
\unskip\
\newblock
\APACrefYearMonthDay{2011}{}{}.
\newblock
{\BBOQ}\APACrefatitle {Mixture cure model with random effects for clustered
  interval-censored survival} {Mixture cure model with random effects for
  clustered interval-censored survival}.{\BBCQ}
\newblock
\APACjournalVolNumPages{Statistics in Medicine}{30}{9}{995--1006}.
\PrintBackRefs{\CurrentBib}

\bibitem [\protect \citeauthoryear {%
Xie%
\ \BBA {} Yu%
}{%
Xie%
\ \BBA {} Yu%
}{%
{\protect \APACyear {2021}}%
}]{%
Xie21a}
\APACinsertmetastar {%
Xie21a}%
\begin{APACrefauthors}%
Xie, Y.%
\BCBT {}\ \BBA {} Yu, Z.%
\end{APACrefauthors}%
\unskip\
\newblock
\APACrefYearMonthDay{2021}{}{}.
\newblock
{\BBOQ}\APACrefatitle {Promotion time cure rate model with a neural network
  estimated nonparametric component} {Promotion time cure rate model with a
  neural network estimated nonparametric component}.{\BBCQ}
\newblock
\APACjournalVolNumPages{Statistics in Medicine}{40}{15}{3516--3532}.
\PrintBackRefs{\CurrentBib}

\bibitem [\protect \citeauthoryear {%
Yakovlev%
\ \BBA {} Tsodikov%
}{%
Yakovlev%
\ \BBA {} Tsodikov%
}{%
{\protect \APACyear {1996}}%
}]{%
Yak96}
\APACinsertmetastar {%
Yak96}%
\begin{APACrefauthors}%
Yakovlev, A\BPBI Y.%
\BCBT {}\ \BBA {} Tsodikov, A\BPBI D.%
\end{APACrefauthors}%
\unskip\
\newblock
\APACrefYear{1996}.
\newblock
\APACrefbtitle {Stochastic Models of Tumor Latency and their Biostatistical
  Applications} {Stochastic models of tumor latency and their biostatistical
  applications}.
\newblock
\APACaddressPublisher{Singapore}{World Scientific}.
\PrintBackRefs{\CurrentBib}

\bibitem [\protect \citeauthoryear {%
Yin%
}{%
Yin%
}{%
{\protect \APACyear {2005}}%
}]{%
Yin05b}
\APACinsertmetastar {%
Yin05b}%
\begin{APACrefauthors}%
Yin, G.%
\end{APACrefauthors}%
\unskip\
\newblock
\APACrefYearMonthDay{2005}{}{}.
\newblock
{\BBOQ}\APACrefatitle {{B}ayesian cure rate frailty models with application to
  a root canal therapy study} {{B}ayesian cure rate frailty models with
  application to a root canal therapy study}.{\BBCQ}
\newblock
\APACjournalVolNumPages{Biometrics}{61}{2}{552--558}.
\PrintBackRefs{\CurrentBib}

\bibitem [\protect \citeauthoryear {%
Yin%
\ \BBA {} Ibrahim%
}{%
Yin%
\ \BBA {} Ibrahim%
}{%
{\protect \APACyear {2005}}%
{\protect \APACexlab {{\protect \BCnt {1}}}}}]{%
Yin05}
\APACinsertmetastar {%
Yin05}%
\begin{APACrefauthors}%
Yin, G.%
\BCBT {}\ \BBA {} Ibrahim, J\BPBI G.%
\end{APACrefauthors}%
\unskip\
\newblock
\APACrefYearMonthDay{2005{\protect \BCnt {1}}}{}{}.
\newblock
{\BBOQ}\APACrefatitle {Cure rate models: a unified approach} {Cure rate models:
  a unified approach}.{\BBCQ}
\newblock
\APACjournalVolNumPages{Canadian Journal of Statistics}{33}{4}{559--570}.
\PrintBackRefs{\CurrentBib}

\bibitem [\protect \citeauthoryear {%
Yin%
\ \BBA {} Ibrahim%
}{%
Yin%
\ \BBA {} Ibrahim%
}{%
{\protect \APACyear {2005}}%
{\protect \APACexlab {{\protect \BCnt {2}}}}}]{%
YinIb05}
\APACinsertmetastar {%
YinIb05}%
\begin{APACrefauthors}%
Yin, G.%
\BCBT {}\ \BBA {} Ibrahim, J\BPBI G.%
\end{APACrefauthors}%
\unskip\
\newblock
\APACrefYearMonthDay{2005{\protect \BCnt {2}}}{}{}.
\newblock
{\BBOQ}\APACrefatitle {A general class of {B}ayesian survival models with zero
  and nonzero cure fractions} {A general class of {B}ayesian survival models
  with zero and nonzero cure fractions}.{\BBCQ}
\newblock
\APACjournalVolNumPages{Biometrics}{61}{}{403--412}.
\PrintBackRefs{\CurrentBib}

\bibitem [\protect \citeauthoryear {%
Zeng%
, Yin%
\BCBL {}\ \BBA {} Ibrahim%
}{%
Zeng%
\ \protect \BOthers {.}}{%
{\protect \APACyear {2006}}%
}]{%
Zen06}
\APACinsertmetastar {%
Zen06}%
\begin{APACrefauthors}%
Zeng, D.%
, Yin, G.%
\BCBL {}\ \BBA {} Ibrahim, J\BPBI G.%
\end{APACrefauthors}%
\unskip\
\newblock
\APACrefYearMonthDay{2006}{}{}.
\newblock
{\BBOQ}\APACrefatitle {Semiparametric transformation models for survival data
  with a cure fraction} {Semiparametric transformation models for survival data
  with a cure fraction}.{\BBCQ}
\newblock
\APACjournalVolNumPages{Journal of the American Statistical
  Association}{101}{474}{670--684}.
\PrintBackRefs{\CurrentBib}

\end{thebibliography}

\end{document}